\documentclass[11pt,a4paper]{article}
\pdfoutput=1

\usepackage{bm}
\usepackage{amsmath,amssymb}
\usepackage{cite}

\usepackage{graphicx}
\usepackage{color}
\usepackage{upgreek}

\usepackage{hyperref} %%put this package declaration last to make work since it redefines things!!%%
\hypersetup{colorlinks=true, citecolor=blue, filecolor=black, linkcolor=blue, urlcolor=blue, pdfpagemode=UseNone}
\numberwithin{figure}{section}
\numberwithin{equation}{section}

\newcommand{\be}{\begin{equation}}
\newcommand{\ee}{\end{equation}}
\newcommand{\bea}{\begin{eqnarray}}
\newcommand{\eea}{\end{eqnarray}}
\def\beal#1\eeal{\begin{align}#1\end{align}}   %one & only for aligning
\def\besp#1\eesp{\begin{multline}#1\end{multline}} %split an equation with first line left aligned & later right aligned

\newcommand{\TRM}[1]{#1}
\usepackage[normalem]{ulem}

\newcommand{\cL}{\mathcal{L}}
\newcommand{\cO}[1]{\mathcal{O}_{#1}}
\newcommand{\pO}[1]{\mathcal{O}^{(#1)}}
\newcommand{\ph}{\varphi}
\newcommand{\vp}{\varphi}
\newcommand{\tp}{\tilde{\vp}}
\newcommand{\vpi}{\uppi}
\newcommand{\tpi}{\tilde{\vpi}}
\newcommand{\tg}{\tilde{g}}
\newcommand{\tV}{\tilde{V}}
\newcommand{\tK}{\tilde{K}}
\newcommand{\Lz}{ {\Lambda_0} }
\newcommand{\Lzp}[1]{\Lambda_0^{#1}}
\newcommand{\Lp}{a\Lambda_\p}
\newcommand{\p}{\mathrm{p}}
\newcommand{\dd}[2]{\delta_{\!\phantom{(} #1}^{\!(#2)}\!(\ph)}
\newcommand{\ddp}[3]{\delta_{\!\phantom{(} #1}^{\!(#2)}\!(#3)}
\newcommand{\Lm}[1]{\mathfrak{L}_{#1}}
\newcommand{\Lmm}{\Lm-}

\newcommand{\tc}[2]{\tilde{c}^{#2}_{#1}}
\newcommand{\cc}[2]{\mathring{c}^{#2}_{#1}}
\newcommand{\Sh}{\mathcal{S}}
\newcommand{\I}{\mathcal{I}}
\newcommand{\mon}{\sigma} %{\mathfrak{m}}
\newcommand{\fV}{\mathcal{V}}
\newcommand{\tfV}{\tilde{\mathcal{V}}}
\newcommand{\M}{\mathcal{M}}

\newcommand{\h}{h }
\newcommand{\htot}{H}
\newcommand{\hmn}{\h_{\mu\nu}}

\newcommand\ie{\textit{i.e.}\ }
\newcommand\eg{\textit{e.g.}\ }
\newcommand\cf{\textit{cf.}\ }

\newcommand{\aka}{{a.k.a.}\ }

\newcommand{\viz}{{\it viz.}\ }
\newcommand{\half}{\tfrac{1}{2}}

\begin{document}

\begin{titlepage}
%\begin{flushright}
%%{\tt hep-ph/yymmnn}
%{\tt SHEP xx-xx}
%\end{flushright}

\begin{center}
{\huge \bf Renormalization group properties in the conformal sector: towards perturbatively renormalizable quantum gravity}

%of the conformal sector in quantum gravity}
%Relevant directions for the conformal factor in perturbative quantum gravity 
%\emph{or: Through the conformal factor to(wards) perturbatively renormalizable quantum gravity }} 
%\vskip.3cm
%{\huge \bf  and etc} 
\end{center}
\vskip1cm

%\title{xxx}
%\author{Tim R. Morris}

\begin{center}
{\bf Tim R. Morris}
\end{center}

%\affiliation{
\begin{center}
{\it STAG Research Centre \& Department of Physics and Astronomy,\\  University of Southampton,
Highfield, Southampton, SO17 1BJ, U.K.}\\
\vspace*{0.3cm}
{\tt  T.R.Morris@soton.ac.uk}
\end{center}

\abstract{The Wilsonian renormalization group (RG) requires Euclidean signature.
% where %With this signature, 
The conformal factor of the metric then has a wrong-sign kinetic term, which has a profound effect on its RG properties.
%Changing the sign of the kinetic term in scalar field theory has a profound effect on its Wilsonian renormalization group properties. 
Generically for the conformal sector, complete flows exist only in the reverse direction  (\ie from the infrared to the ultraviolet).  %After imposing a quantisation conditon on the large field decay of the bare interactions, t
The Gaussian fixed point supports infinite sequences of composite eigenoperators of increasing infrared relevancy (increasingly negative mass dimension), which are orthonormal and complete for bare interactions that are square integrable under the appropriate measure. 
%a quantisation condition, that bare interactions decay sufficiently fast for large field, makes th Restriction to this space is achieved 
These eigenoperators are non-perturbative in $\hbar$ and evanescent. 
%(\ie vanish when the ultraviolet regulator is removed). 
For $\mathbb{R}^4$ spacetime, each renormalized physical operator exists but only has support at vanishing field amplitude. In the generic case of infinitely many non-vanishing couplings, if a complete RG flow exists, it is characterised in the infrared by a scale $\Lambda_\p>0$, beyond which the field amplitude is exponentially suppressed. On other spacetimes, %characterised by some 
of length scale $L$, the flow ceases to exist once a certain universal measure of inhomogeneity exceeds $O(1)+2\pi L^2\Lambda^2_\p$. Importantly for cosmology, the minimum size of the universe is %then set by 
thus tied to the degree of inhomogeneity, with spacetimes of vanishing size being required to be almost homogeneous. We initiate a study of this exotic quantum field theory at the interacting level, 
and discuss what the full theory of quantum gravity should look like, %: 
one 
which must thus be
%The full theory of quantum gravity will be
perturbatively renormalizable in Newton's constant but non-perturbative %ly quantum.} % 
in $\hbar$.}

%at vanishing scales, or in %a compact spacetime 
%other spacetimes, 
%the physical field amplitude is in general only exponentially suppressed, in the latter case by a universal function of the spacetime geometry. \TRM{uniformity, cosmology}
%We initiate the study of this exotic quantum field theory at the interacting level.
%Since the conformal factor of the metric has such a wrong sign kinetic term, {\it a priori} the existence of these operators opens a route to (a) 
%%perturbatively renormalizable quantum gravity, 
%renormalizable quantum gravity, perturbative in Newton's constant but non-perturbative in $\hbar$,
%(with \emph{possibly} %possessing 
%novel and welcome phenomenological properties). We describe our %so-far-unsuccessful attempt 
%attempts to construct such a theory.\emph{or: We make the first tentative steps towards such a theory. OR We begin its construction.}}

\end{titlepage}

\tableofcontents

\newpage

\section{Introduction}
\label{sec:Intro}

If one follows the by--now--standard procedures of perturbative quantum field theory, %such as those that make up the Standard Model of particle physics, 
then one finds that quantum gravity suffers from the problem that it is not perturbatively renormalizable. The natural coupling constant is $\kappa=2/M$, where $M$ is the reduced Planck mass. In terms of Newton's gravitational constant $G$, we have $\kappa^2 = 32\pi G$. Given that $\kappa$ has negative mass dimension, perturbative non-renormalizability is expected already from simple power counting arguments. Kinematic accidents allow pure gravity at one loop to be free of divergences \cite{tHooft:1974toh} (after a reparametrisation of the metric $g_{\mu\nu}$), but with generic matter or at two loops, no such miracle occurs \cite{tHooft:1974toh,Goroff:1985sz,Goroff:1985th,vandeVen:1991gw}.

We will show however, %without going outside the structure of perturbative 
that within quantum gravity, perturbative in $\kappa$ and starting from the (kinetic parts of the) Einstein Hilbert action,\footnote{This is thus not related to asymptotic safety \cite{Weinberg:1980,Reuter:1996},
although we will draw on some insight from that field.}  there exists a distinguished set of composite operators, dependent on the conformal factor of the metric and non-perturbative in $\hbar$, that are promising for a route 
%\textit{a priori} would appear to offer a way 
out of this dead end. Even at the linearised level, \ie for vanishingly small coupling(s), they have novel infrared properties which have the potential to explain long-standing puzzles in cosmology, and black holes, and maybe even lead to experimentally measurable quantum gravity effects, as discussed later in the introduction and in secs. \ref{sec:compact-linear} and \ref{sec:QG}.

% which present even at vanishingly small $\kappa$, which provide a completely new perspective quantum gravity in general and , as we discuss later.
% 
%\TRM{\textit{do explain some things... and not just through these IR properties}
%which have the potential to explain long-standing puzzles in cosmology (dark energy, inflation), shed new light on black holes, and maybe even explain conflicting experimental measurements of $G$ \cite{Mohr:2015ccw}.}
% % for Newton's gravitational constant $G$ \cite{•}.

%Over the years many inventive ideas have been put forward to circumvent this problem. They all have in common that 
To understand clearly why there is this possibility, we will need to work with the deeper understanding of renormalization afforded by the Wilsonian RG (renormalization group) \cite{Wilson:1973,Morris:1998}. Since an essential ingredient in this framework is the quasi-local effective action constructed from integrating out fluctuations at short distances, 
%we require $|x-y|\to0\implies x\to y$, and so
we will need to work with a Euclidean signature metric.\footnote{so that indeed for two points $x$ and $y$, $|x-y|\to0\implies x\to y$.} Then one meets the infamous problem that the Euclidean Einstein-Hilbert action,\footnote{Our conventions are $R_{\mu\nu}=R^\alpha_{\ \mu\alpha\nu}$, and $[\nabla_\mu,\nabla_\nu]v^\lambda = R_{\mu\nu\phantom{\lambda}\sigma}^{\phantom{\mu\nu}\lambda}v^\sigma$.}
\be 
\label{EH}
S_{EH} =  \int\!\! d^4x \, \cL_{EH}\,,\qquad \cL_{EH} = -2\sqrt{g} R/\kappa^2\,,
\ee
is unbounded from below, 
%\footnote{We set $\kappa^2 = 32\pi G$. Then $\kappa=2/M$, where $M$ is the reduced Planck mass.} 
so that the Euclidean partition function 
\be 
\label{Z}
\mathcal{Z} = \int\!\! \mathcal{D}g_{\mu\nu}\ {\rm e}^{-S_{EH}}
\ee
will fail to converge. Expanding the metric about flat space as
\be 
\label{h}
g_{\mu\nu} = \delta_{\mu\nu} + \kappa\, \htot_{\mu\nu}\,,
\ee
we have 
\be 
\label{EHbilinear}
\cL_{EH} = \frac12 \left(\partial_\lambda \htot_{\mu\nu}\right)^2 -2 \left(\partial_\lambda \ph\right)^2 - \left(\partial^\mu \htot_{\mu\nu}\right)^2 +2\,\partial^\alpha\! \ph\, \partial^\beta \htot_{\alpha\beta} + O(\htot)^3\,,
\ee
where contraction is with the background metric $\delta_{\mu\nu}$, and we have defined 
$\ph = \half %h^{\tot\,\mu}_\mu. 
H^{\,\mu}_\mu$.
Adding a Feynman -- De Donder gauge fixing term 
\be 
\label{Feynman-DeDonder}
\left(\partial^\alpha \htot_{\alpha\beta} -\partial_\beta \ph\right)^2
\ee 
and splitting the fluctuation field into its SO$(4)$ irreducible parts
\be 
\label{ph-traceful}
\htot_{\mu\nu} = \h_{\mu\nu} + \half \delta_{\mu\nu} \ph
\ee
(so $\h_{\mu\nu}$ is traceless),
the problem is clearly visible in the wrong sign kinetic term for $\ph$:
\be 
\label{Gaussian}
\cL^{\rm kinetic}_{EH} = \frac12 \left(\partial_\lambda \h_{\mu\nu}\right)^2 -\frac12 \left(\partial_\lambda\ph \right)^2\,.
\ee
Since the metric is now expressed as
\be 
\label{param-perturbative}
g_{\mu\nu} = \delta_{\mu\nu} \left( 1+\frac{\kappa}{2}\,\ph \right) +\kappa \,\h_{\mu\nu}\,,
\ee
we see that $\ph$ is the perturbation that leads to an overall local rescaling of the metric. It is called the conformal factor, or the dilaton (even though it is not a separate field here but part of the metric).
%for which the conformal factor perturbation, $\ph$, has the wrong sign.
The authors of ref. \cite{Gibbons:1978ac} proposed to fix the problem by continuing the conformal factor functional integral along the imaginary axis: $\ph\mapsto i\ph$. Instead, we will %find another way of coping with 
keep this ``conformal factor instability'',  and find another way of coping, which moreover has a clear physical motivation. Indeed it seems that the conformal factor instability is the key that opens the door to formulating continuum quantum gravity.

Mathematically, the first step is to recast \eqref{Z} into differential form by using an exact %renormalization group 
RG equation for the corresponding effective action. Then there is no immediate difficulty in solving for the latter \cite{Reuter:1996}.
Within this Wilsonian framework, the problem with perturbative renormalizability is simply that the interactions 
\be 
\label{irrelevant}
\sim \htot^n \partial \htot \partial \htot \qquad (n\ge1) 
\ee
form \emph{irrelevant} operators (of dimension $n+4$). This follows by na\"\i ve scaling arguments which are
nevertheless correct at the Gaussian fixed point \eqref{Gaussian}.
Such interactions cannot therefore build a continuum field theory around the Gaussian fixed point, since a continuum field theory requires operators corresponding to (marginally) relevant directions. Of course this only repackages the power counting arguments, although if taken as gospel it already implies that miraculous cancellations of divergences were never a way out. 

But why rule out non-polynomial interactions? As we will review in the next section, for theories with the right sign kinetic term, the polynomial interactions form a complete orthonormal set of eigenoperators (operators with a well defined scaling dimension). 
%under the appropriate measure. 
Non-polynomial perturbations with definite scaling dimension at finite field, do not scale correctly at large field. They do not emanate from the Gaussian fixed point and after RG evolution to the IR (infrared), they can be re-expanded in terms of the polynomial perturbations and thus do not lead to new continuum physics \cite{Morris:1996nx,Morris:1996xq,Bridle:2016nsu}. 

When we change the sign of the kinetic term, this conclusion changes radically. The same arguments that ruled out non-polynomial interactions for ordinary scalar field theory now imply that the eigenoperator spectrum degenerates, and even includes a continuous component \cite{Dietz:2016gzg}. Completeness and orthonormality properties are lost. Furthermore the Wilsonian RG now naturally flows in the reverse direction, meaning that generic flows to the infrared fail at some finite cutoff scale \cite{Bonanno:2012dg,Dietz:2016gzg}.

Now we add just one, albeit crucial, observation. As part of the definition of quantization, we are free to impose that bare interactions are exponentially decaying for large $\ph$ (see sec. \ref{sec:quantisation}). 
Stated more precisely, we require them to be square integrable over amplitude $\ph\in (-\infty,\infty)$ with weight 
\be 
\label{weight}
\exp \left(\ph^2/2\Omega_\Lambda\right)\,,
\ee
where $\Omega_\Lambda= |\langle \ph(x) \ph(x) \rangle |$ is the (magnitude of the) free propagator at coincident points, regularised by a UV (ultraviolet) cutoff $\Lambda$. 
%we will be more precise in the next section) 
Then as we will see, the eigenoperator spectrum is again discrete, complete, and orthonormal. 

%As we will see in sec. \ref{sec:minus},
Working within the conformal sector (\ie retaining only $\ph$), the rest of the properties of this remarkable quantum field theory follow ineluctably. 
We will see that the eigenoperators are non-perturbative in $\hbar$, and are evanescent \cite{Bollini:1973wu} \ie vanish when the ultraviolet regulator is removed. In $\mathbb{R}^4$, the physical (renormalized) operators become proportional to ($\ph$-derivatives of) $\delta(\ph)$. On other spacetimes, the physical operators are instead exponentially decaying with the amplitude decay scale related to $1/L$, where $L$ is a typical length scale in the manifold. However if the manifold is sufficiently inhomogeneous, in the sense of inducing more than an $O(1)$ change to a certain universal finite size effect (see sec. \ref{sec:compact}), each operator individually ceases to exist because the flow to the infrared ends prematurely.

%thus individually, presumably  switch off any physically measurable fluctuations. 
Infinitely many of the eigenoperators are relevant. They therefore can be used to build a non-trivial continuum limit about the Gaussian fixed point, in other words a perturbatively renormalizable quantum field theory. In the case that an infinite number of these relevant couplings are non-vanishing, which is inevitable beyond first order perturbations, new effects emerge.
%with however an infinite number of (relevant) couplings. 
In fact even at the linearised level, when an infinite number of these relevant couplings are non-vanishing, it typically happens that at some lower scale $\Lambda\sim\Lambda_\p>0$, %$\Lambda=a\Lambda_\p$ ($a$ is a non-universal number) 
the expansion over eigenoperators no longer converges. The result can nevertheless be resummed by transforming to field conjugate momentum space.
%the corresponding effective interactions fail to be square integrable under \eqref{weight}. In the appropriate norm  and 
As we will show, convergence fails  either because the RG flow itself ceases to exist, or because the interactions are no longer square integrable under \eqref{weight} but instead have exponential decay set by $\Lambda_\p$, which we therefore recognise as an
%instead exponentially suppressed by some 
\emph{amplitude suppression scale}. %$\Lambda_\p>0$ (in turn determined by the couplings).

Now on other manifolds the flow exists only if the inhomogeneity remains smaller than the $O(1)$ correction plus $2\pi L^2\Lambda^2_\p$. As already mentioned in the Abstract, this property is clearly significant for the theory of cosmology, but also surely for black holes and more generally (see secs. \ref{sec:compact-linear} and \ref{sec:QG}). The fact that such dramatic behaviour is already evident at the linearised level, \ie even at vanishing overall coupling, suggests that such quantum gravity effects could be experimentally measurable. However confirming this will require understanding the dynamics, which in turn requires the full development of the quantum gravity, \ie not just the conformal sector.

Indeed a further significant step is to embed this structure into gravity, where we need also to maintain a quantum version of diffeomorphism invariance at the renormalized level. We discuss the issues in sec. \ref{sec:QG}. Although the conformal sector has an infinite number of renormalized couplings, these get subsumed effectively into the parametrisation of the metric. As we will see, renormalizability of the diffeomorphism invariant local operators is controlled by one particular eigenoperator, which turns out to have just the right dimension to rule in the Einstein-Hilbert term and rule out all the higher derivative terms. The wrong sign kinetic term makes the scalar theory non-unitary (see sec. \ref{sec:unitarity}) but this problem will not affect gravity when continued back to Minkowski signature, where only the two transverse traceless modes actually propagate and the conformal mode is not dynamical. Since the quantum field theory is built around the Gaussian fixed point, it will be perturbatively renormalizable, in particular in $\kappa$. Although the theoretical structure is so constraining that General Relativity is guaranteed to be the  low energy effective classical description, since the eigenoperators in the conformal sector are non-perturbative in $\hbar$, and indeed vanish in the limit that $\hbar\to0$, in reality the  theory of gravity will be non-perturbatively quantum and have no classical limit, no matter how small $\kappa$ is taken to be.\footnote{unless $\kappa$ is set to zero, in which case we are left with only free gravitons}
%Certain not necessarily tiny effects, such as those related to inhomogeneity as outlined above, will lie outside a classical description. 
\bigskip

The structure of the rest of the paper is as follows. Until the final two sections we will be almost exclusively concerned with the conformal sector considered on its own. In Euclidean flat space, this is just a single component scalar field theory with the wrong sign kinetic term. The significance of this change in sign for the Wilsonian RG about the Gaussian fixed point, can only be properly understood once the standard case with positive kinetic term is thoroughly understood. Therefore in the next section (sec. \ref{sec:plus}) we review the latter case. In sec. \ref{sec:minus} we change the sign of the kinetic term and develop the consequences for the Wilsonian RG, working in flat Euclidean $\mathbb{R}^4$ spacetime  and with linearised perturbations. With the example of the potential, we see  in sec. \ref{sec:non-deriv} that typical flows for the RG exist only in the reverse direction and that the eigenspectrum degenerates. We show that one sequence of perturbations has however a Hilbert space structure. In sec. \ref{sec:quantisation} we define the bare interactions to lie in this space as part of the definition of quantisation. As intimated earlier, everything else follows as a logical consequence. In particular we develop the properties of these eigenoperators, which for the potential are all relevant, and introduce $\Lambda_\p$ which (up to a non-universal constant) marks the infrared scale where the expansion over eigenoperators breaks down. In sec. \ref{sec:general} we see that for entire flows, $\Lambda_\p$ is a physical quantity, namely the \emph{amplitude suppression scale}. In sec. \ref{sec:examples}, we illustrate with a simple representative example. In sec. \ref{sec:derivative-ops}, we derive the form of the general eigenoperator \ie containing also space-time derivatives. In sec. \ref{sec:perturbation-th}, we start the development of the full non-linear theory. In sec. \ref{sec:unitarity}, we highlight the physical flaws that such a scalar field theory has, if considered in its own right. As already addressed above, these problems are not expected to be inherited by a full theory of quantum gravity. In sec. \ref{sec:QG} we consider what form this latter theory must take (and the phenomenological consequences). However first in sec. \ref{sec:compact} we examine the behaviour of RG flows on a manifold other than $\mathbb{R}^4$. There we see that $\Lambda_\p$ has another dramatic r\^ole to play, limiting the degree of inhomogeneity according to the size of the universe. 
%Finally, in sec. \ref{sec:conclusions} we summarise and draw some further conclusions.

\section{Scalar field theory with positive kinetic term}
\label{sec:plus}

In this section we review the RG structure of scalar field theory about the Gaussian fixed point, establishing that the eigenoperator spectrum is given by a complete set of orthonormal polynomial interactions.  In particular we explain why non-polynomial interactions that satisfy the eigenoperator equation, do not behave correctly in the UV (ultraviolet) and after RG evolution to the IR (infrared) can be re-expanded in terms of the polynomial interactions.
%in particular after RG evolution (towards the IR) of the corresponding perturbation, the non-polynomial solutions can be expanded in terms of these. 
This was analysed in great detail in ref. \cite{Bridle:2016nsu}, see also 
\cite{Morris:1996nx,Morris:1996xq}, however the focus there was different and model approximations were used (in particular the so-called Local Potential Approximation). Here, and in the rest of this paper, we make no approximations beyond the use of perturbation theory where it is legitimate to do so.

Not only do we need to work in Euclidean signature (as already remarked in the Abstract and the beginning of the Introduction) but we also need to work on $\mathbb{R}^4$, since for fixed points to exist, the space-time itself should look exactly the same at all scales. Momentum is therefore a useful concept. These remarks may seem trivial but it is important to underline these points for when we adapt this framework to gravitation.
%will be important later when we apply these ideas to gravitation. 

After integrating out high momentum modes, we can rewrite the partition function exactly in terms of a Wilsonian effective action \cite{Wilson:1973,Morris:1993}
\be 
\label{total-Wilsonian}
S^{\mathrm{tot},\Lambda}[\ph] = S^\Lambda[\ph] + \frac{1}{2}\ph\cdot (\Delta^{\Lambda})^{\!-1}\!\!\cdot \ph\,,
\ee
where 
\be 
\label{DeltaUV}
\Delta^\Lambda(p) := \frac{C^\Lambda(p)}{p^2}
\ee
is here the massless propagator regularised by some smooth ultraviolet cutoff profile $C^\Lambda(p)\equiv C(p^2/\Lambda^2)$. Later, when we change the sign of the propagator, we will still define $\Delta^\Lambda$ to be \eqref{DeltaUV}, \ie positive as displayed above. Qualitatively, for $|p|<\Lambda$, $C^\Lambda(p)\approx1$ and mostly leaves the modes unaffected, while for $|p|>\Lambda$ its r\^ole is to suppress modes.
We require that $C(p^2/\Lambda^2)$ is a monotonically decreasing function of its argument, that $C^\Lambda(p) \to 1$ for $|p|/\Lambda\to0$, and for $|p|/\Lambda\to\infty$, $C^\Lambda(p) \to0$  sufficiently fast to ensure that all momentum integrals are regulated in the ultraviolet.

After discarding a field independent part, the interactions satisfy the Wilson/Polchinski flow equation \cite{Polchinski:1983gv,Morris:1993}
	\begin{equation}
	\label{pol+}
	\frac{\partial}{\partial \Lambda}S^\Lambda[\ph]=\frac{1}{2}\frac{\delta S^\Lambda}{\delta\ph}\cdot \frac{\partial\Delta^\Lambda}{\partial \Lambda}\cdot			\frac{\delta S^\Lambda}{\delta\ph}-\frac{1}{2}\text{tr}\bigg[\frac{\partial\Delta^\Lambda}{\partial \Lambda}\cdot \frac{\delta^{2}S^\Lambda}			{\delta\ph\delta\ph}\bigg]\,.
	\end{equation}
The first term on the right hand side encodes the tree level corrections, while the second term encodes the quantum corrections. Had we carried $\hbar$, it would appear in front of this latter term.
 We want the quasi-local solutions of this equation, \ie solutions $S^\Lambda$ that can be written as the space-time integral of a Lagrangian, which in turn can be written as an (infinite) expansion in space-time derivatives of $\ph$. Such solutions correspond to a local Kadanoff blocking and exist if $C^\Lambda$ is smooth. 

The Gaussian fixed point is the trivial solution $S^\Lambda[\ph]=0$. To find the eigenoperators we linearise around the fixed point:
\begin{equation}
	\label{d-pol}
	\frac{\partial}{\partial \Lambda}\,\delta S^\Lambda[\ph]=-\frac{1}{2}\,\text{tr}\bigg[\frac{\partial\Delta^\Lambda}{\partial \Lambda}\cdot \frac{\delta^{2} }{\delta\ph\delta\ph}\bigg] \delta S^\Lambda[\vp]\,.
	\end{equation}
Let us first consider non-derivative interactions. Thus we write:
\be 
\label{linearised-v}
\delta S^\Lambda = \epsilon\! \int\!\!d^4x\, V\!\left(\ph(x),\Lambda\right)\,,
\ee
where $\epsilon$ is taken small enough to justify the linearised approximation. The Wilsonian RG consists of a Kadanoff blocking followed by a rescaling back to the original size. This second step is conveniently incorporated by using
scale independent variables formed from the dimensionless combinations using $\Lambda$:
\be 
\label{scale}
x^\mu = \tilde{x}^\mu/\Lambda\,,\qquad\ph = \Lambda\, \tilde{\ph}\,,\qquad V = \Lambda^4\, \tilde{V}\,, \qquad t = \ln(\mu/\Lambda)\,.
\ee
We have noted that at the Gaussian fixed point the scaling dimension of $\ph$ is its engineering dimension. We have also defined the so-called RG time $t$ to increase in the direction of course graining, as in ref. \cite{Wilson:1973}, and introduced the usual arbitrary finite energy scale $\mu$.
Eigenoperators are then operators with well defined scaling dimension $4 -\lambda$, when expressed in these variables, which thus take the form
\be 
\label{lambda-v}
\tilde{V}(\tilde{\ph},t) = \left(\frac{\mu}{\Lambda}\right)^\lambda \tilde{V}(\tilde{\ph})\,,
\ee
the prefactor being the RG evolution of the scaled coupling $\tg_\lambda=\epsilon\, {\rm e}^{\lambda t}$
at linearised order, the associated dimensionful coupling thus being 
\be 
\label{dimensionful-g-lambda}
g_\lambda= \epsilon\mu^\lambda\,. 
\ee
Such operators are relevant if $\lambda>0$, marginal if $\lambda=0$, and irrelevant if $\lambda<0$. The continuum limit is constructed by giving non-vanishing values for the couplings 
associated to relevant and marginally relevant directions since these shoot out of the fixed point as $\Lambda$ is lowered from $\Lambda=\infty$ (\ie $\tg_\lambda\to0$ as $t\to-\infty$), and also to any strictly marginal couplings. The continuum limit is parametrised by these couplings, and characterised by the resulting ``RG trajectory'' as $\Lambda$ is lowered. The (marginally) irrelevant couplings do not survive as separate parameters in the continuum limit since they lead to trajectories that fall back into the fixed point, rather they parametrise the basin of attraction of the fixed point \cite{Wilson:1973,Morris:1998}.

Although we will mostly restrict ourselves to this linear regime in the current paper, to be precise and to set the context let us briefly sketch the complete construction. Since the (marginally) relevant couplings increase as $\Lambda$ is lowered, we need to handle the full non-linear exact RG. Then we need to define what we still mean by such $\tg_\lambda(\Lambda)$ in the non-linear regime, which we can do conveniently by imposing some renormalization conditions on $S^\Lambda$. (Such a renormalization condition is also needed for the kinetic term and leads to rescaling the field, \ie wavefunction renormalization.)
The dimensionful $g_\lambda(\Lambda)$ will then run with scale once we enter the non-linear regime. Since as described in the previous paragraph, the asymptotic UV behaviour for these couplings provides the boundary conditions that completely fixes the flow, solutions on the RG trajectory can be written in self-similar form as $S^\Lambda = S(\tg_\lambda)$, \ie   where $\Lambda$ dependence only enters through the dimensionless (marginally) relevant couplings. Substituting this form back into the flow equation, the corresponding $\beta_\lambda$ functions can be read off from the renormalization conditions. 
Choosing finite values for the couplings at a finite scale $\Lambda$, and integrating up these $\beta$ functions, thus solves for the full RG trajectory. To the extent that there is something to prove, it is only that one should establish that there exist such solutions that match into the asymptotic UV regime. Since the $\tg_\lambda(\Lambda)$, or equivalently $g_\lambda(\Lambda)$, are  finite at finite scales they are \textit{de facto} renormalized couplings. Since renormalization is in this sense automatic, we will not tend to use this terminology. On the other hand, we should distinguish these from the finitely related \emph{physical} couplings. We will define these later via the Legendre effective action.

Returning to the linear regime we will mostly treat in this paper, we note that since each dimensionful coupling then does not run, its `bare' value in the far UV and the `renormalized' value in the IR, both coincide with \eqref{dimensionful-g-lambda}. We can and will also choose a physical renormalization condition so that  \eqref{dimensionful-g-lambda} coincides with the physical coupling.

%by shooting out of the fixed point along the 
%relevant and marginally relevant directions  as $\Lambda$ is lowered from $\Lambda=\infty$, \ie by choosing non-vanishing conjugate couplings which satisfy 
%Choosing a solution such that the $\tg_\lambda(\Lambda)$ are finite when $\Lambda$ reaches physical scales, means that these couplings now play the 
%When $\Lambda$ is finite, the continuum limit is parametrised by 
% the physical couplings associated to these operators (we will define the physical couplings later) and characterised by the resulting ``RG trajectory'' as $\Lambda$ is lowered. 

From \eqref{pol+}, the eigenoperator equation is thus
\be 
\label{eigen+}
-\lambda\, \tilde{V}(\tp) -\tilde{\ph}\, \tilde{V}' + 4\, \tilde{V} = -\frac{\tilde{V}''}{2{a}^2}\,,
\ee
where a prime is differentiation with respect to the field argument, and we have defined the  dimensionless one-loop massless tadpole integral\footnote{Although $a$ is a pure number, it is non-universal, clearly dependent on the cutoff profile.}
\be 
\label{a}
\frac1{2{a}^2} = \frac1{2\Lambda}\, \frac{\partial}{\partial\Lambda}\, \Omega_\Lambda = 
\int\!\frac{d^4\tilde{p}}{(2\pi)^4}\, \frac{C(\tilde{p}^2)}{\tilde{p}^2}\,,
\ee
taking $a>0$, and $\Omega_\Lambda= \Lambda^2/2a^2$ is the dimensionful version:
\be 
\label{Omega}
\Omega_\Lambda :=  |\langle \ph(x) \ph(x) \rangle | = \int\!\frac{d^4p}{(2\pi)^4}\, \Delta^\Lambda(p)\,.
\ee
We have %noted that 
defined it as the magnitude of the  propagator evaluated at a point. Here the propagator is positive anyway, but later it won't be. 

\TRM{Equation \eqref{eigen+} is of Sturm-Liouville type. Its
quantised solutions are in fact} the Hermite polynomials 
\be 
\label{On}
\cO{n}(\tp) = H_n(a\tp)/(2a)^n = \tp^n -n(n-1)\tp^{n-2}/4a^2 +\cdots\,,
\ee
with $\lambda = 4-n$ and $n$ a non-negative integer. The (scaling) dimension of the operator $\cO{n}$ is thus $4-\lambda=n$, coinciding with the engineering dimension $[\vp^n]$. The lower powers in \eqref{On} are there to correct for operator mixing as $\Lambda$ is varied and appear with increasing powers of $\hbar$. 
%we see that it appears in the combination $\hbar/a^2$: we have a classical piece plus corrections polynomial in $\hbar$. 
They arise from tadpole corrections, which are the only quantum corrections remaining at linearised order. 
%In particular the negative mass term correction in the marginal operator $\cO{4} = \tp^4-3\tp^2/a^2+3/4a^4$ is there to cancel exactly the quadratic mass term divergence, thus automatically giving the renormalized $\vp^4$ interaction (at linearised level) in the continuum limit. 

As is well known, for a marginal operator we need to go beyond linearised order to decide its fate. And once we go beyond linearised order, $\cO{4}$ becomes (marginally) irrelevant. For a true continuum limit, the only relevant directions (and thus renormalized couplings) in this case are therefore the mass term $\cO{2}$ and the vacuum energy $\cO{0}$ (which however without gravity carries no physics), so that we are left with a massive free theory, a somewhat inconvenient conclusion for illustrating the general structure -- but we trust the latter will be sufficiently clear despite these specific facts.
%makes four dimensional scalar field theory somewhat inconvenient for illustrating the general structure, but we can ignore this in the ensuing since we need only linearised order.

\TRM{From the general Sturm-Liouville theory we know that the} $\cO{n}$ form an orthonormal set:
\be
\label{orthonormal+}
\int^\infty_{-\infty}\!\!\!\! d\tp\,\, {\rm e}^{-a^2\tp^2} \cO{n}(\tp) \cO{m}(\tp) = \frac{1}{a}\left(\frac{1}{2a^2}\right)^n\! n!\sqrt{\pi}\,\delta_{nm}\,,
\ee
which is complete in $\Lm+$, 
% for any perturbation $\tV(\tp)\in\Lm+$, where $\Lm+$ is 
the natural space for Wilsonian interactions around a positive kinetic energy term. This Hilbert space 
is the
space of functions that are square integrable under the \TRM{Sturm-Liouville} measure ${\rm e}^{-a^2\tp^2}$.  By all this we mean that if $\tV(\tp)\in\Lm+$, and we set
\be 
\label{tg+}
\tg_n = \frac{a}{\sqrt{\pi}}\frac{(2a^2)^n}{n!}\int^\infty_{-\infty}\!\!\!\! d\tp\,\, {\rm e}^{-a^2\tp^2} \cO{n}(\tp) \tV(\tp)\,,
\ee
the norm-squared of the remainder vanishes as we extend to an infinite series, \ie
\be 
\label{completeness-proof}
\int^\infty_{-\infty}\!\!\!\! d\tp\,\, {\rm e}^{-a^2\tp^2} \left( \tV(\tp) - \sum_{n=0}^N \tg_n \cO{n}(\tp)\right)^{\!2}\to0\quad{\rm as}\quad N\to\infty\,. 
\ee
%Clearly, at the linearised level the RG evolution of $\tV(\tp)$ is then determined. 
In this sense, all perturbations in $\Lm+$ are 
%that are square integrable under ${\rm e}^{-a^2\tp^2}$, they are 
described %in terms of the infinite number of 
by a countable infinity of couplings $\tg_n$, and their RG evolution is just given by the RG evolution of these couplings. 

To form the bare action at $\Lambda=\Lambda_0$, which we can take to be the initial condition for the flow equation \eqref{pol+}, 
we need to choose the bare couplings $\tg^{(\lambda)}_0\equiv\tg^{(\lambda)}(\Lz)$. The simplest choice is to set the bare irrelevant couplings to zero. A more general choice that stays within the basin of attraction of the Gaussian fixed point (at least in perturbation theory) is to set %the $\tg^{(\lambda)}_0\equiv\tg^{(\lambda)}(\Lz)$ 
them to some finite fixed value \ie to set $g^{(\lambda)}_0 = \tg^{(\lambda)}_0 \Lambda_0^\lambda$, where $\tg^{(\lambda)}_0$ is a fixed pure number if $\lambda<0$. In contrast the bare relevant couplings $\tg^{(\lambda)}_0$ need to follow the flow and thus vanish as $\Lz\to\infty$.
%are set so that the renormalized couplings will be finite: 
At the linearised level,
$\tg^{(\lambda)}_0 = g^{(\lambda)} \Lambda_0^{-\lambda}$ where now $g^{(\lambda)}$ is some fixed finite dimension-$\lambda$ coupling (the renormalized coupling) if $\lambda>0$. Note that as $\Lambda_0\to\infty$ in order to form the continuum limit, the linearised approximation for the relevant  couplings becomes ever more valid at scales close to the bare scale.

The effective action \eqref{total-Wilsonian} can in this way provide the bare action, and studying its evolution away from the bare action provides us with direct access to the Wilsonian RG framework,
but does not directly furnish us with physical quantities.  We can access these latter in a useful way by replacing the cutoff $C^\Lambda$ in \eqref{DeltaUV} by 
\be 
\label{sum-rule}
C^\Lz_k(p) = C^\Lz(p) - C^k(p)\,,
\ee
thus the theory is now also infrared regulated at scale $k$ \cite{Morris:2015oca}. Then writing the Legendre effective action as 
\be
\label{total-Gamma}
\Gamma^{\text{tot},\, \Lz}_{k}[\ph]= \Gamma^\Lz_{k}[\ph]+
	\frac{1}{2}\varphi\cdot \left(\Delta_{k}^\Lz\right)^{\!-1}\!\!\cdot \varphi\,,
\ee
where
\be 
\label{sum-ruled-Delta}
\Delta^\Lz_k = \Delta^\Lz - \Delta^k\,,
\ee
we have the identity (up to discarding a field independent part on the right hand side)
\be 
\Gamma^\Lz_\Lz[\ph] = S^\Lz[\ph]\,, 
\ee
which provides us with the initial condition for a flow with respect to the infrared cutoff, %which reads in this notation
the latter taking the form \cite{Nicoll1977,Morris:1993,Morris:2015oca} (see also \cite{Bonini:1992vh,Wetterich:1992}):
\begin{equation}
	\label{Gamma.flow+}
	\frac{\partial}{\partial k}\Gamma^\Lz_{k}[\varphi]=-\frac{1}{2}\text{tr}\bigg[\bigg(1+\Delta^\Lz_k\cdot 						\frac{\delta^{2}\Gamma^\Lz_{k}}{\delta\varphi\delta\varphi}\bigg)^{\!-1}\frac{1}									{\Delta^\Lz_k}\frac{\partial\Delta^\Lz_k}{\partial k}\bigg]\,.
	\end{equation}
At the Gaussian fixed point the Legendre effective action has just the field independent part $\Gamma^\Lz_k[\ph]=-\half\,\text{tr}\ln\Delta^\Lz_k$. Once again looking at linearised perturbations, we have:
\begin{equation}
	\label{d-Gamma.flow+}
	\frac{\partial}{\partial k}\, \delta\Gamma^\Lz_{k}[\varphi]=-\frac{1}{2}\,\text{tr}\bigg[\frac{\partial\Delta^k}{\partial k}\cdot \frac{\delta^{2} }{\delta\ph\delta\ph}\bigg] \delta \Gamma^\Lz_{k}[\varphi]\,,
	\end{equation}
where we have used \eqref{sum-rule} to simplify the expression.  We see that $\delta \Gamma^\Lz_{k}[\varphi]$ satisfies an identical equation to \eqref{d-pol} with $k$ now playing the r\^ole of a UV cutoff. The reason for this is as follows. Since at the linearised level the flow equation has become insensitive to the overall UV cutoff $\Lz$, we can send this to infinity. Then we can note that $\Gamma_\Lambda:=\Gamma_\Lambda^\infty$ is related to $S^\Lambda$ by a Legendre transform:  $\Gamma_\Lambda$ carries the purely quantum, 1PI (one particle irreducible), parts of $S^\Lambda$ \cite{Morris:1993,Morris:1998,Morris:2015oca}.\footnote{See also \cite{Keller:1990ej,Bonini:1992vh}. The existence of $\Lambda\to\infty$ flows is a different matter, and is why in general such a complete (renormalized) trajectory must terminate at a fixed point.} However at the linearised level there are only quantum corrections and thus the flow equations coincide. Setting
\be 
\label{linearised-v-G}
\delta \Gamma^\Lz_k[\ph] = \epsilon\! \int\!\!d^4x\, V\left(\ph(x),k\right)\,,
\ee
the interaction potential will therefore satisfy the same eigenoperator equation \eqref{eigen+} as that for the Wilsonian effective action, only with $\Lambda$ replaced by $k$ in \eqref{scale} and \eqref{lambda-v}.

Now suppose that we add $g_n\pO{n}_\Lz(\vp)$ to the bare action \ie at $k=\Lambda=\Lz$. By this we mean that we add in scaled units $\tg_n \cO{n}(\tp)$, where $\tg_n = \tg_n(\Lz)=g_n /\Lambda_0^{4-n}$. To linearised order, and in scaled units, this evolves in a self-similar way by construction, \ie keeps the same form, with the dimensionless variables formed using the appropriate scale. In particular we recognise that the coupling becomes
\be 
\label{mult-evolve}
 \left(\frac{\Lz}{k}\right)^{4-n}\!\!\tg_n(\Lz) = \frac{g_n}{\ k^{4-n}}= \tg_n(k) \,.
% \left(\frac{\Lz}{k}\right)^{4-n}\!\!\tg_n(\Lz)\, \cO{n}(\tp)\,.
\ee
Therefore, using \eqref{scale}, the dimensionful (unscaled) interaction is
\be 
\label{physical-Onk}
g_n\,\pO{n}_k(\vp) =k^4  \frac{g_n}{\ k^{4-n}} \,\,\cO{n}\!\left({\vp}/{k}\right) 
%= g_n k^n \cO{n}\!\left({\vp}/{k}\right)  
=
g_n\left( \vp^n -n(n-1) \frac{k^2}{4a^2}\vp^{n-2}+\cdots\right)\,,
\ee
\ie 
\be 
\label{physical-OnL}
\pO{n}_\Lambda(\vp)= \Lambda^n\,\cO{n}(\vp/\Lambda) = \vp^n -n(n-1) \frac{\Lambda^2}{4a^2}\vp^{n-2}+\cdots\,.
\ee
Again we note that in the Wilsonian RG framework, the operator and associated coupling are already the renormalized ones once the cutoff $k$ falls to physical scales. In addition in the limit $k\to0$, we find the universal \emph{physical} interaction, as it appears in the Legendre effective action. In this case we thus find $\pO{n}(\vp) :=\lim_{k\to0} \pO{n}_k(\vp)$, where:
%removing the IR cutoff, \ie taking the limit $k\to0$, leaves just the now--universal interaction
\be 
\label{physical-On}
g_n \pO{n}(\vp) =  g_n\vp^n\,.
\ee 
%in the Legendre effective action. 
Recalling the discussions above, we see that for relevant directions this is finite and $g_n$ indeed corresponds to the  physical coupling, while for the irrelevant directions $g_n$ is proportional to an inverse power of $\Lz$ and thus tends to zero in the continuum limit $\Lz\to\infty$.
%\be 
%k^4 \left(\frac

%If we start in the bare action with a general effective potential $\tV_\Lz(\tp)\in\Lm+$, then expressing it in terms of the couplings $\tg_n$, the statement that it is in $\Lm+$ means that the following sum converges:
%\be 
%\int^\infty_{-\infty}\!\!\!\! d\tp\,\, {\rm e}^{-a^2\tp^2} \tV^2_\Lz(\tp) = \frac{\pi}{a^2}\sum_{n=0}^\infty \frac{n!}{(4a^4)^n}\, \tg_n^2 \ <\infty\,.
%\ee
%If we can treat the potential at the linearised level, then from the multiplicative evolution \eqref{mult-evolve}, we have at scale $k$:
%\be 
%\int^\infty_{-\infty}\!\!\!\! d\tp\,\, {\rm e}^{-a^2\tp^2} \tV^2_k(\tp) = \frac{\pi}{a^2}\sum_{n=0}^\infty \frac{n!}{(4a^4)^n}\, \tg_n^2 \left(k/\Lz\right)^{2n-4}\,.
%\ee
%We see that for all $0<k<\Lz$, the convergence of the series is only improved, and therefore the evolved $\tV_k(\tp)$ remains in $\Lm+$.

Note that Wilsonian RG properties are only manifest in scaled variables. For example the statement that relevant perturbations emanate from the Gaussian fixed point in the ultraviolet, \ie vanish as $\Lambda\to\infty$, is only true in scaled variables. In dimensionful terms the tadpole correction terms actually diverge in this limit, as can be seen from \eqref{physical-OnL}. In particular for example, the negative mass term correction in the marginal operator $\cO{4} = \tp^4-3\tp^2/a^2+3/4a^4$, which is fixed and finite in scaled variables, is there to cancel exactly the quadratic mass term divergence (the divergence responsible for the naturalness problem in Higgs physics), thus automatically giving the renormalized $\vp^4$ interaction (at linearised level) in the continuum limit as we saw above.

The evolution \eqref{mult-evolve} can be understood in this way more conventionally in terms of Feynman diagrams. We will make that connection clearer later for the novel operators we discover  for scalar field theory with wrong sign kinetic term. Similarly we could continue the development by including (spacetime) derivative interactions, and also in going beyond linearised order into perturbation theory with the (marginally) relevant couplings. Of course we are only rephrasing standard knowledge here, so instead  we make these developments directly for the novel operators in sec. \ref{sec:minus}. 

Now we address the fate of non-polynomial solutions to \eqref{eigen+}, which cannot be understood purely in terms of Feynman diagrams since non-perturbative physics is required (although of a rather trivial sort). At first sight the general solution of \eqref{eigen+}, which can be written in terms of Kummer functions, allows for new eigenoperators, in particular ones for which $\lambda>0$ and which thus can be used to build exotic continuum limits \cite{HHOrig}. Their large field behaviour grows as $\sim \tp^{\lambda-5}\exp (a^2\tp^2)$, so they lie outside $\Lm+$. However it is not true that these solutions provide new continuum limits \cite{Morris:1996nx,Morris:1996xq,Bridle:2016nsu}. The reason is that for fixed $\epsilon$, no matter how small, the linearised approximation, \eqref{linearised-v} or \eqref{linearised-v-G},
is not valid for large field. To find the correct evolution for such a perturbation, one needs to use the full non-linear flow equation in the large field regime. Thus such solutions will also evolve differently depending on whether we regard this as a perturbation that is purely quantum or includes the classical corrections  \cite{Bridle:2016nsu}. The simplest picture arises from taking it to be purely quantum. In fact since $\Gamma_\Lambda$ diverges at large field, it follows from \eqref{Gamma.flow+} that the right hand side vanishes and thus the dimensionful (unscaled) interaction does not evolve at all in this limit. Correspondingly in scaled units the interaction will follow ``mean field evolution''. Adding such an operator to the bare $\Gamma_\Lz$, we thus find at any other scale $\Lambda$, in the large field regime $\tp\gg \Lz/(\Lambda\sqrt{\ln\epsilon})$,
\be 
\sim \epsilon\, \tp^{\lambda-5} \left(\frac{\Lambda}{\Lz}\right)^{\lambda-1}\!\!\!\exp\left\{ a^2\tp^2 \Lambda^2/\Lzp2\right\}\,.
\ee
To be a relevant perturbation we want this scaled version to vanish as $\Lambda\to\infty$ so that we return to the Gaussian fixed point in this limit, but we see that actually the scaled perturbation diverges in this limit. On the other hand for RG evolution into the IR, once $\Lambda<\Lz/\sqrt{2}$, the interaction is inside $\Lm+$ and thus can be expanded as a convergent series in terms of the $\cO{n}$. 

Actually, also when we add the perturbation $\tg_n \cO{n}$ to the bare $\Gamma_\Lz$, the linearised approximation is not valid for large field for $n>2$. Mean field evolution therefore takes over here too, and thus at scale $\Lambda$ it becomes 
\be 
 \left(\frac{\Lz}{\Lambda}\right)^{4}\!\! \tg_n(\Lz)\,\cO{n}(\tp \Lambda/\Lz) \,.
\ee
The difference is that at large field this just gives us back self-similar evolution and \eqref{mult-evolve} \cite{Morris:1996nx,Morris:1996xq,Bridle:2016nsu}.

At the same time these observations establish that a general (not necessarily small) 1PI perturbation $\tV_\Lz(\tp)$ that starts in $\Lm+$, remains in $\Lm+$ under evolution to the IR, %as far as its large field behaviour is concerned, 
and thus the complete evolution can be understood in terms of the corresponding $\tg_n(k)$. However note that  $\Lm+$ is not defined when the cutoff reaches $k=0$. In the limit $k\to0$, the relevant interactions diverge, so $\tV_k(\tp)$ is itself ill defined in this limit. This can be seen in \eqref{mult-evolve}, although
of course the linearised approximation breaks down before this happens. Nevertheless the mass and vacuum energy terms clearly will in general diverge in scaled units using $k$ (see also \eg \cite{Bridle:2016nsu}). For these reasons the property $\tV_k(\tp)\in\Lm+$ can only be defined for all $\Lz\ge k>0$ (\ie excluding the limit $k\to0$).

%Lowering $\Lambda$ achieves the Kadanoff blocking step of the Wilsonian RG. The rescaling back to the original size is straightforwardly incorporated by forming dimensionless variables using $\Lambda$. We should use the scaling dimension of the field, however around the Gaussian fixed point that coincides with the engineering dimension. 

\section{Scalar field theory with negative kinetic term}
\label{sec:minus}

Now we change the sign of the kinetic term. At face value this makes no sense, since now  the functional integral in the partition function no longer even na\"\i vely converges, while the momentum cutoff profile, instead of exponentially suppressing the integrand, makes matters worse. 
But gravity presents us with this problem if we are to understand it in Wilsonian terms, since then we must consider fluctuations about Euclidean $\mathbb{R}^4$ (\cf  beginning sec. \ref{sec:plus}). Therefore we need to generalise what we mean by quantum field theory in this case in order to make progress. Instead of following ref. \cite{Gibbons:1978ac} and analytically continuing so as to remove the sign, we keep the sign and
seek an appropriate generalisation of the structure outlined in the previous section. 

%Therefore we replace 
We begin by replacing \eqref{total-Wilsonian} and \eqref{total-Gamma}  by\footnote{Note that for convenience $\Delta^\Lambda$ in \eqref{DeltaUV}, \cf also \eqref{a} and \eqref{Omega}, are defined to be positive.}
\be 
\label{total-}
S^{\mathrm{tot},\Lambda}[\ph] = S^\Lambda[\ph] - \frac{1}{2}\ph\cdot (\Delta^{\Lambda})^{\!-1}\!\!\cdot \ph\,,\qquad \Gamma^{\text{tot},\, \Lz}_{k}[\ph]= \Gamma^\Lz_{k}[\ph]-
	\frac{1}{2}\varphi\cdot \left(\Delta_{k}^\Lz\right)^{\!-1}\!\!\cdot \varphi\,.
\ee
As a result, $\Delta\mapsto -\Delta$ in the flow equations \eqref{pol+},
%\footnote{However, in preparation for later we write $C_\Lambda \equiv C^\infty_\Lambda = 1 - C^\Lambda$, using \eqref{sum-rule}, and also $\dot{}\equiv \partial_t$, \cf \eqref{scale}.}
%\eqref{d-pol}, 
\eqref{Gamma.flow+} and 
\eqref{d-Gamma.flow+}:\footnote{In preparation for later we have reinstated $\Delta^\Lz_k$ in the last equation.}
	\bea
	\label{pol-}
	\frac{\partial}{\partial\Lambda} S^\Lambda[\ph] &=&{-}
	\frac{1}{2}\,\frac{\delta S^\Lambda}{\delta\ph}\cdot \frac{\partial\Delta^\Lambda}{\partial\Lambda}\cdot			\frac{\delta S^\Lambda}{\delta\ph}+\frac{1}{2}\,\text{tr}\bigg[\frac{\partial\Delta^\Lambda}{\partial\Lambda}\cdot \frac{\delta^{2}S^\Lambda}			{\delta\ph\delta\ph}\bigg]\,, \\
%	\end{equation}
%\begin{equation}
	\label{Gamma.flow-}
	\frac{\partial}{\partial k}\Gamma^\Lz_{k}[\varphi] &=& -\frac{1}{2}\,\text{tr}\bigg[\bigg(1-\Delta^\Lz_k\cdot 						\frac{\delta^{2}\Gamma^\Lz_{k}}{\delta\varphi\delta\varphi}\bigg)^{\!-1}\frac{1}{\Delta^\Lz_k}\frac{\partial\Delta^\Lz_k}{\partial k}\bigg]\,,\\
%	\end{equation}
%\begin{equation}
	\label{d-Gamma.flow-}
	\frac{\partial}{\partial k}\, \delta\Gamma^\Lz_{k}[\varphi] &=&{-}
	\frac{1}{2}\,\text{tr}\bigg[\frac{\partial\Delta^\Lz_k}{\partial k}\cdot \frac{\delta^{2} }{\delta\ph\delta\ph}\bigg] \delta \Gamma^\Lz_{k}[\varphi]\,.
	\eea
%To save space we will refer to these equations, but keep in mind that they now have these sign corrections. 
%Let us pause to note that t
This makes these equations backward-parabolic, which means in particular that the Cauchy initial value problem for flow towards the IR is not well posed. To elucidate this and further consequences, we will again begin by considering non-derivative interactions at the linearised level.

\subsection{Non-derivative eigenoperators}
\label{sec:non-deriv}

The linearised flow for the potential 
\be 
\label{flow-V}
\partial_t V(\vp,t) = -\Omega_\Lambda\, %\partial_\vp^2 
V''(\vp,t)\,,
\ee 
can be written:
\be 
\label{heat}
\frac{\partial}{\partial T}\, V(\vp,T) = \frac1{4a^2} V''(\vp,T)\,,
\ee
which is now in the form of the heat diffusion equation,  with a `time' $T=\Lambda^2$, which runs towards the UV.
This means that for a general `initial' potential $V(\vp,T_0)$, well-defined flows only exist towards the UV (which is thus also an issue for the full flow equations \cite{Bonanno:2012dg,Dietz:2016gzg}). In the other direction, the bare action must be chosen carefully if the flow is to exist all the way to $k\to0$. Indeed, this is already intuitively clear from the connection to heat diffusion. Flowing in the UV  direction, the potential will diffuse out, becoming ever smoother. On the contrary, flows towards the IR will reverse the diffusion process, typically resulting in a $V(\vp,T)$ that develops singularities in $\vp$ at some critical `time' $T=T_\p :=a^2\Lambda^2_\p$, after which the flow ceases to exist, \ie the flow typically ends at some $k =\Lp>0$.\,\footnote{Although we do not address the asymptotic safety scenario in this paper, since the flow is again backward-parabolic, it is clear that generic flows towards the IR, will end at some critical scale there also \cite{Bonanno:2012dg,Dietz:2016gzg}.} (We include the factor $a$ in the definition of $\Lambda_\p$ for convenience: as we will see in sec. \ref{sec:general}, in other circumstances $\Lambda_\p$ can then have a universal meaning.)
%For a generic bare action, the flow will end at a singular point at some finite scale $k =\Lambda_\p>0$, as we will see. 

The fact that flow is more naturally in the reverse direction suggests that 
%This does not mean that flows do not exist from the UV to the IR, but rather leads to the expectation that 
universality should be found in the UV limit rather than the IR. Indeed we are
%It is thus already a signal of what we are 
about to find that the Gaussian fixed point now supports eigenoperators of arbitrarily high relevancy (\ie for RG time reversed flows, playing the r\^ole of the usual hierarchy of irrelevant operators).

In fact without further restriction, the situation is worse than that. 
%\be 
%-\partial_t V(\vp,t) = \Omega_\Lambda\, %\partial_\vp^2 
%V''(\vp,t)\,,
%\ee
%which is indeed effectively the heat equation
%but with the time running backwards.\footnote{Writing $\Omega_\Lambda=\Lambda^2/2a^2$, `thermal' time is identified with $\Lambda^2$.}
To realise the Wilsonian RG, we need to use the scaled variables \eqref{scale}, giving
\be 
\label{scaled-flow-V}
\Lambda\frac{\partial}{\partial \Lambda}\tilde{V}_\Lambda(\tp) -\tilde{\ph}\, \tilde{V}'_\Lambda(\tp) +4\, \tilde{V}_\Lambda(\tp) = {\tilde{V}''_\Lambda}(\tp)/{(2{a}^2)}\,.
\ee
Then setting $\tV_\Lambda(\tp)=\, {\rm e}^{\lambda t}\, \tV(\tp)$,
we get the eigenoperator equation \eqref{eigen+} except with a plus sign on the right hand side:
\be 
\label{eigen-}
-\lambda\, \tilde{V}(\tp) -\tilde{\ph}\, \tilde{V}' + 4\, \tilde{V} = \frac{\tilde{V}''}{2{a}^2}\,.
\ee
The change in relative sign between the $\tp \tilde{V}'$ and $\tilde{V}''$ term means that at large field one no longer has exponentially growing solutions. Instead they behave at worst as 
\be 
\label{GaussianEigenasymptotic}
\tV\propto\tp^{4-\lambda}+\frac{(4-\lambda)(3-\lambda)}{4a^2}\tp^{2-\lambda}+
{O}(\tp^{-\lambda})\,,
\ee
which is generically an asymptotic series which is also subject to exponentially decaying corrections 
$\sim\tp^{\lambda-5} \,{\rm e}^{-a^2\tp^2}$. %\exp (-a^2\tp^2)$. 
For $\lambda>2$, such solutions justify linearisation of the right hand side of \eqref{Gamma.flow-} ever more accurately as $\tp\to\infty$ and thus are not ruled out by the large field analysis reviewed in sec. \ref{sec:plus}, while for $\lambda\le2$ mean field analysis still allows these perturbations since it just gives back the correct multiplicative evolution \ie $(\Lz/k)^\lambda \tV$. Thus the large field test rules out none of the solutions \cite{Dietz:2016gzg}.

These solutions divide into three sets as follows \cite{Dietz:2016gzg}. For every $\lambda$ there are two linearly independent solutions, an odd and even Kummer function, which thus form a continuous eigenoperator spectrum. For $\lambda$ not an integer, by adjustment of their ratio, one can arrange for zero coefficient for the asymptotic series in \eqref{GaussianEigenasymptotic} on one side $\tp\to\pm\infty$, leaving behind the exponentially decaying corrections, but on the other side $\tp\to\mp\infty$ it will then have \eqref{GaussianEigenasymptotic} as its asymptotic behaviour. %(see \eg \cite{Mohammedi2016}). 
At $\lambda$ an integer, one of the two Kummer functions degenerates, thus forming two discrete spectra: at $\lambda=4-n$ there are the polynomial solutions, which now read $\cO{n}(\tp)=H_n(ia\tp)/(2ia)^n$; for $\lambda=5+n$, we have an infinite tower of exponentially decaying `super-relevant' eigen-operators:%\footnote{Corresponding solutions existed also for \eqref{eigen+} but were exponentially growing and thus by the large field analysis did not evolve correctly. The first expression can be found by taking the Fourier transform of \eqref{eigen-}, while the second follows from substituting $\tV\mapsto \tV\, {\rm e}^{-a^2\tp^2}$ and comparing to \eqref{eigen+}.}
\be 
\label{delta}
\delta_n(\tp) := \frac{a}{\sqrt{\pi}} \frac{\partial^n}{\partial\tp^n} \, {\rm e}^{-a^2\tp^2} =  \frac{a}{\sqrt{\pi}} (-a)^n H_n(a\tp) \, {\rm e}^{-a^2\tp^2}
\,,\qquad \lambda=5+n\,,  %[154.9], [155.1]
\ee
$n$ a non-negative integer, whose dimension is thus 
\be 
\label{delta-dim}
[\delta_n]=4-\lambda = -1-n\,.
\ee 
Solutions corresponding to these latter also existed for \eqref{eigen+} but were exponentially growing and thus by the large field analysis did not evolve correctly. 

The second expression in \eqref{delta} follows from substituting $\tV\mapsto \tV\, {\rm e}^{-a^2\tp^2}$ into \eqref{eigen-} and comparing to \eqref{eigen+}. The first expression can be found by substituting the Fourier transform:
\be 
\label{Fourier}
\tV(\tp) = \int^\infty_{-\infty}\! \frac{d\tpi}{2\pi}\, \tfV(\tpi)\, \mathrm{e}^{i\tpi\tp}\,,
\ee
%into \eqref{eigen-}, 
where $\tpi=\vpi\Lambda$ is the scaled conjugate momentum, giving the general solution:
\be 
\label{general-sol}
\tfV(\tpi) = (i\tpi)^{\lambda-5} \exp\left(-\frac{\tpi^2}{4a^2}\right)\,.
\ee
This has power-law asymptotics \eqref{GaussianEigenasymptotic}, generated by the singularity at $\tpi=0$, except that the singularity is absent when $\lambda=5+n$ where it gives \eqref{delta}.
% \eqref{eigen-}, giving:
%\be 
%(5-\lambda) \tV(\tpi) +\tpi \frac{\partial}{\partial\tpi} = -\frac{\tpi^2}{2a^2}\tV
%\ee

\TRM{Equation \eqref{eigen-} is still of Sturm-Liouville type, but the} Sturm-Liouville weight function is now ${\rm e}^{+a^2\tp^2}$. 
Defining $\Lmm$ to be the space of square integrable functions under this measure, 
the polynomials and the continuous spectrum of Kummer functions lie outside this space. 
%are not square integrable under this measure. 
However the exponentially decaying solutions  %are,
lie inside $\Lmm$
and indeed %then 
form a complete orthonormal basis for this Hilbert space:
\be
\label{orthonormal-}
\int^\infty_{-\infty}\!\!\!\! d\tp\,\, {\rm e}^{a^2\tp^2} \delta_n(\tp)\, \delta_m(\tp) = \frac{a}{\sqrt{\pi}}\left({2a^2}\right)^n\! n!\,\delta_{nm}\,,
\ee
(where we used the 2$^{\rm nd}$ eqn in \eqref{delta}) so that if $\tV(\tp) \in\Lmm$ and
%is square integrable under ${\rm e}^{+a^2\tp^2}$, and
\be 
\label{tg}
\tg_n = %\frac{\sqrt{\pi}}{a}\frac{1}{(2a^2)^nn!}
\frac{\sqrt{\pi}}{2^na^{2n+1}n!}
\int^\infty_{-\infty}\!\!\!\! d\tp\,\, {\rm e}^{a^2\tp^2} \delta_n(\tp)\, \tV(\tp)\,,
\ee
the norm-squared of the remainder vanishes as we extend to an infinite series, \ie
\be 
\label{completeness-proof-}
\int^\infty_{-\infty}\!\!\!\! d\tp\,\, {\rm e}^{a^2\tp^2} \left( \tV(\tp) - \sum_{n=0}^N \tg_n\, \delta_n(\tp)\right)^{\!2}\to0\quad{\rm as}\quad N\to\infty\,. 
\ee
This structure is %almost 
the generalisation we are looking for. 
%In this sense, if $\tV_\Lambda(\tp)\in\Lmm$ then
%\be 
%\label{tv-Lambda}
%\tV_\Lambda(\tp) = \sum_{n=0}^\infty \tg_n\, \delta_n(\tp)\,.
%\ee
%By Fourier transform we have \eqref{fourier-sol} with \eqref{fourier-expansion}, where $g_n = \tg_n\, \Lambda^{5+n}$.

\subsection{Quantisation condition}
\label{sec:quantisation}

Although we cannot exclude the solutions outside $\Lm-$ by their large field RG properties, we can exclude them by fiat. We thus choose, \emph{as part of the definition of quantisation}, to insist that the bare interactions must lie in $\Lmm$. 

If we consider a finite sum of the basis operators \eqref{delta} then this quantisation condition is clearly respected by the RG at the linear level, since the operators evolve multiplicatively. Indeed if at the bare scale $\Lambda=\Lz$, $\delta_n(\tp)$ appears linearly with a sufficiently small coupling 
$
\tg_n = %\tg_n(\Lz)= 
g_n /\Lz^{5+n}
$, 
then at some other scale it will still take this form but with $\tg_n=g_n/\Lambda^{5+n}$ (where $g_n$ is held fixed).

If an infinite number of couplings are switched on, then by our quantisation condition we require:
\be 
\label{tv-bare}
\tV_{\Lz}(\tp) = \sum_{n=0}^\infty \tg_n\, \delta_n(\tp)\  \in \Lmm\,.
\ee
Again, if $\tV$ is small enough to trust the linear RG evolution, then at another scale $\tV_\Lambda(\tp)$ takes the same form with $\Lz$ replaced by $\Lambda$ (\ie both explicitly, and implicitly in the scaled quantities): %\TRM{Utilising the Fourier transform we can express this as \eqref{fourier-sol}, where $\fV_\p$ is given by \eqref{fourier-expansion}.}
\be 
\label{tv-evolved}
\tV_{\Lambda}(\tp) = \sum_{n=0}^\infty \tg_n\, \delta_n(\tp)\,.
\ee
Using \eqref{orthonormal-},  we can compute the norm-squared of the evolved potential:
\be 
\label{tv-norm-squared}
\int^\infty_{-\infty}\!\!\!\! d\tp\,\, {\rm e}^{a^2\tp^2} \tV_\Lambda^2(\tp) = 
\frac{a}{\Lambda^{10}\sqrt{\pi}}  \sum_{n=0}^\infty n!\, g_n^2 \left(\frac{2a^2}{\Lambda^2}\right)^{\!n}\,.
\ee
By \eqref{tv-bare}, the series on the right hand side converges for $\Lambda=\Lambda_0$.
We thus see that $\tV_\Lambda(\tp)\in\Lmm$ and remains small 
for all $\Lambda\ge\Lz$. This is why we interpret the quantisation condition $\tV_\Lambda(\tp)\in\Lmm$ as operating at the bare level. Since all the couplings $g_n$ are relevant, we set them to be finite at physical scales, whence they parametrise the most general RG trajectory. The above properties ensure that the Wilsonian effective interaction continues to satisfy the quantisation condition as $\Lambda\to\infty$. Indeed $\tV_\Lambda(\tp)\to0$ in this limit, \ie it emanates from the Gaussian fixed point, as it should to describe the RG trajectory. Like any continuum limit, it can be regarded conceptually as existing in its own right,  without the need to postulate  a microscopic theory. However if we do entertain that possibility, then the quantisation condition provides a hint as to the form this microscopic theory would have to take. 

On the other hand the generic case will be that the $g_n$ are such that the series \eqref{tv-norm-squared} has a finite radius of convergence $1/\Lambda=1/(\Lp)$ where, by \eqref{tv-bare}, $\Lp\le\Lz$. Then $\tV_\Lambda(\tp)\notin\Lmm$ for all $\Lambda<\Lp$, %less than this critical scale. 
although also generically as $\Lambda$ decreases, the linearised approximation breaks down. 
In any case  once $\tV_\Lambda(\tp)\notin\Lmm$, the expansion over the basis \eqref{delta} no longer converges. There are two possible reasons for $\tV_\Lambda(\tp)$ exiting $\Lmm$: either $\tV_\Lambda(\tp)$ itself has developed divergences, or it grows too fast for large $\tp$ so that the integral in \eqref{tv-norm-squared} no longer converges for $\tp\to\pm\infty$. 
In the former case the flow ceases to exist, as we anticipated earlier by using the heat equation. We will see an explicit example later. In the latter case 
its evolution can still be described by the appropriate flow equation, namely \eqref{scaled-flow-V}, more generally \eqref{pol-} or \eqref{Gamma.flow-}. Since the flow is first order in $\Lambda$, it can be uniquely determined by supplying as boundary condition the expansion over the basis, for any $\Lambda>\Lp$. 
%where we still have $\tV_\Lambda(\tp)\in\Lmm$. %(and similarly for \eqref{Gamma.flow-}). 
%However as we will see, it is the former case that supplies the reason for exiting $\Lmm$: $\tV_\Lambda(\tp)$ develops singularities, after which it ceases to exist.\footnote{Although we do not address the asymptotic safety scenario in this paper, since the flow is again backward-parabolic, it is clear that generic flows will end at some critical $\Lp$ there also.} This can already be understood intuitively from the heat equation  \eqref{heat}: flowing in the UV direction, a localised $V(\vp,T)$ will diffuse out becoming ever smoother. On the contrary, flows towards the IR will reverse the diffusion process, typically ending in some distributional $V(\vp,T)$ at some critical $T=T_\p=\Lambda^2_\p$.
At a formal level, we can still write $\tV_\Lambda(\tp)$ as an expansion over the basis, even for $\Lambda<\Lp$. Indeed at the linearised level it will continue to be \eqref{tv-evolved}, since each term separately satisfies \eqref{scaled-flow-V}. However in this region we need a prescription for resumming the series. We will see that this is provided by working in conjugate momentum space.

The eigenoperators have novel physical properties. Analogously to \eqref{physical-OnL}, we identify the dimensionful bare operator $\dd{\Lz}n$
as the conjugate to the dimension $5\!+\!n$ unscaled coupling $g_n$ in the bare action. Thus, either directly from its dimension \eqref{delta-dim} or by re-expressing the coupling and using \eqref{scale},
% at the linearised level,
\be 
\label{def-physical}
\dd{\Lz}n = %\frac{\Lzp4}{\Lzp{5+n}}\,  \delta_n(\vp/\Lz)= 
{ \delta_n(\vp/\Lz)}/{\Lzp{1+n}}\,,
\ee
and hence (using $a = \Lz/\sqrt{2\Omega_\Lz}$):
\be
\label{physical-dnL}
\dd{\Lz}{n} := \frac{\partial^n}{\partial\vp^n}\, \dd{\Lz}{0}\,, \qquad{\rm where}\qquad \dd{\Lz}0 := \frac{1}{\sqrt{2\pi\Omega_\Lz}}\,\exp\left(-\frac{\vp^2}{2\Omega_\Lz}\right)\,.
\ee
If we restore $\hbar$, it multiplies the right hand side of \eqref{eigen+}, similarly \eqref{eigen-} or \eqref{flow-V}, and thus makes its appearance as the combination
$\Omega_\Lz \propto\hbar\, \Lzp2$. We see that the operators are ``evanescent'' \cite{Bollini:1973wu} in the sense that for fixed field $\vp$, the operators vanish as the UV cutoff is removed ($\Lz\to\infty$). They are also non-perturbative in $\hbar$ with a similar functional form in this respect to instanton \cite{Belavin:1975fg,tHooft:1976snw} or renormalon \cite{tHooft:1977xjm} contributions.
%Continuing the development at linear order, recall that the relevant coupling $g_n$ is finite and of mass dimension $5+n$.

By construction, $V=\dd\Lambda{n}$ is a solution of the unscaled flow equation \eqref{flow-V}. A general solution of the linearised RG is the sum of these with constant coefficients $g_n$:
\be 
\label{expand-V}
V(\vp,\Lambda)= \sum_{n=0}^\infty g_n \,\dd\Lambda{n}\,.
\ee
This is nothing but the sum \eqref{tv-evolved} in dimensionful terms (\ie the same except for overall multiplication by $\Lambda^4$). Since by \eqref{tv-bare}, the sum converges for all $\Lambda\ge\Lz$, it follows that even for an infinite number of non-zero couplings, the potential inherits the properties above, \ie it is non-perturbative in $\hbar$, and $V(\vp,\Lambda)\to0$ as $\Lambda\to\infty$, \ie the full potential is evanescent. Note that this property is logically distinct from the `relevancy' property $\tV_\Lambda(\tp)\to0$ in this limit, established below \eqref{tv-norm-squared}, \cf the discussion for normal field theory below \eqref{physical-On}.

Despite the %arguments above that point to this being 
description so far of an  essentially UV structure, there is nevertheless a dramatic imprint on the far IR limit, that is the continuum physics. 
%Analogously to \eqref{physical-Onk}, we identify the physical (unscaled, renormalized) operator $\dd{k}n$
%as the conjugate to the dimension $5+n$ coupling $g_n$ in the Legendre effective action. Thus at the linearised level,
%\be 
%%\label{def-physical}
%\dd{k}n = \frac{k^4}{k^{5+n}}\,  \delta_n(\vp/k)\,,
%\ee
%and hence (using $a = k/\sqrt{2\Omega_k}$):
Since the scaled eigenoperator is form invariant under the linearised RG, the corresponding dimensionful (and automatically renormalized) operator in the IR cutoff Legendre effective action is just
\be
\label{physical-dnk}
\dd{k}{n} = \frac{\partial^n}{\partial\vp^n}\, \dd{k}{0}\,, \qquad{\rm where}\qquad \dd{k}0 = \frac{1}{\sqrt{2\pi\Omega_k}}\,\exp\left(-\frac{\vp^2}{2\Omega_k}\right)\,.
\ee
%To put it more simply, \TRM{!!} $V=\dd\Lambda{n}$ is a solution of the unscaled flow equation \eqref{flow-V}, a general solution being the sum of these with constant coefficients $g_n$.
Removing the IR cutoff gives us the physical  operators %as they appear in the Legendre effective action 
in an $\mathbb{R}^4$ spacetime:
\be 
\label{physical-dn}
\lim_{k\to0} \dd{k}n = \dd{}n \,,
\ee
\ie the $n^{\rm th}$ derivative of the delta-function.\footnote{The unit normalization here explains our choice in \eqref{delta}.}
If we keep only a finite number of couplings 
%and they are small enough to trust the linear RG, 
then since these interactions  have support only on vanishing amplitude, presumably the physics of the renormalized theory is trivial, effectively just a free theory.  This is true in a flat spacetime of infinite extent only when we remove the IR cutoff. In sec. \ref{sec:compact} we will see that on  a homogeneous non-trivial spacetime (with inherent length scales), the amplitude is only suppressed. However once the manifold is sufficiently asymmetric, the physical operator fails to exist because the flow to the IR ends prematurely.

\begin{figure}[ht]
\centering
\includegraphics[scale=0.35]{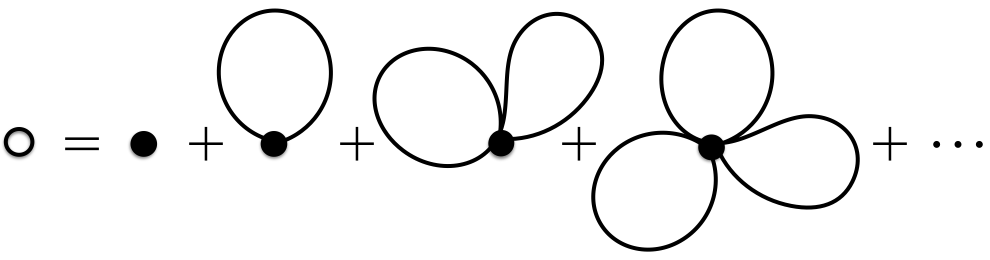}
%\vskip-30pt
\caption{The renormalized eigenoperator is the bare one plus its quantum corrections at linearised level.}
\label{fig:tadpoles}
\end{figure}

The same distributions \eqref{physical-dn} are reached by taking the $\hbar\to0$ limit. In this sense the dynamics is always essentially and non-perturbatively quantum: there is no classical limit.
Let us show how the passage from bare \eqref{physical-dnL} to renormalized \eqref{physical-dnk} can nevertheless be understood in terms of Feynman diagrams.
%At this point it is helpful to see how we flow from the bare interactions 
%\be 
% \dd{\Lz}n = \Lzp{-1-n}\,{\delta_n(\vp/\Lz)}{}\, 
%\ee 
%to the renormalized ones, \eqref{physical-dnk} or \eqref{physical-dn}, directly via a sum of Feynman diagrams. 
%In dimensionful terms, the bare interactions are given by \eqref{def-physical} and thus \eqref{physical-dnk} with $k$ replaced by $\Lz$:
%Recalling the change in sign, and reinstating $\Delta^\Lz_k$, we see that 
The solution to \eqref{d-Gamma.flow-} can be written as:
\be 
\label{tadpoles-evolution}
\int_x \dd{k}n = \exp\left(-\frac{1}{2}\,\text{tr}\left[\Delta^\Lz_k\cdot \frac{\delta^{2} }{\delta\ph\delta\ph}\right]\right) \int_x \dd{\Lz}n\,.
\ee 
The expansion of the exponential gives the expected 1PI Feynman diagrams, as illustrated in fig. \ref{fig:tadpoles}, where the propagator for each tadpole, $-\Delta^\Lz_k$, is defined as in 
 \eqref{sum-ruled-Delta}, and has the sign required from \eqref{total-}. On the other hand  the bare eigenoperator \eqref{physical-dnL} can be written
\be 
\label{bare-delta-Omega}
\dd{\Lz}n = \exp\left(\frac12\Omega_\Lz \frac{\partial^2}{\partial\vp^2}\right) \dd{}n\,,
\ee
as can be seen %by Fourier transforming the Dirac $\delta$-function,
from \eqref{Fourier} and \eqref{general-sol}. Indeed, translating the Fourier transform to unscaled variables using \eqref{def-physical} gives
\be 
\label{dn-Fourier}
\dd{\Lz}n = \int^\infty_{-\infty}\!\! \frac{d\vpi}{2\pi}\, (i\vpi)^n\, \mathrm{e}^{-\frac12\vpi^2\Omega_\Lz+i\vpi\vp}\,,
\ee
after which the result follows by pulling the $\Omega_\Lz$ piece outside the integral. Thus
\be 
\label{bare-delta-prop}
\int_x \dd{\Lz}n = \exp\left(\frac{1}{2}\,\text{tr}\left[\Delta^\Lz\cdot \frac{\delta^{2} }{\delta\ph\delta\ph}\right]\right) \int_x \dd{}n\,.
\ee
Combining this and \eqref{tadpoles-evolution}, and using \eqref{sum-ruled-Delta}, we see that the %left hand side of \eqref{tadpoles-evolution} 
the renormalized operator is given by \eqref{bare-delta-Omega} with $\Lz$ replaced by $k$, and thus by the expression \eqref{physical-dnk}. 

\subsection{General RG flows of the potential at first order in the couplings}
\label{sec:general}

The situation becomes more subtle when an infinite number of couplings are switched on:
as well as solutions that fail to make it to the far IR, there is an infinite dimensional space of  solutions where the physical (\ie  $k=0$) interaction  has support on finite field amplitude.  However if at scale $k$, the (total) interaction lies inside $\Lmm$, we know that, written in dimensionful terms, it must vanish faster than $\exp(-a^2\vp^2/2k^2)/\sqrt{\vp}$ for large $\vp$, which implies that large amplitudes remain significantly damped. In particular if the interaction remains in $\Lmm$ for all $k>0$, then the dimensionful interaction must vanish faster than any such exponential at large $\vp$.
We furnish an example that resolves a puzzle with the form of the physical operators \eqref{physical-dn} at the linear level. The Gaussian fixed point is clearly invariant under the shift of the field by a space-time constant: $\vp(x)\mapsto\vp(x)+\vp_0$. At first sight this symmetry is broken by the operators \eqref{physical-dn}, all of which constrain $\vp$ to zero amplitude. Note that this is not forced by the restriction to be integrable under the measure ${\rm e}^{+a^2\tp^2}$ at the appropriate scales. In fact this breaking is illusory since in the bare action we can add an infinite number of eigenoperators:
\be 
\label{delta-shift+RG}
\tg_m\, \delta_m(\tp+\tp_0) = \tg_m  \sum_{n=0}^\infty \frac{\tp_0^n}{n!}\,\delta_{n+m}(\tp)\,,
\ee 
where, from the first of \eqref{delta}, we have noted that
\be 
\label{d-delta}
%\frac{\partial}{\partial}
\partial_{\tp}\,\delta_n(\tp) = \delta_{n+1}(\tp)\,.
\ee
We see that the corresponding series in \eqref{tv-norm-squared} has an infinite radius of convergence and thus \eqref{delta-shift+RG} remains in $\Lmm$ for all $k>0$. (As with the discussion at the end of sec. \ref{sec:plus}, $k=0$ is excluded.)
Under RG evolution $\delta_{n+m}(\tp)$ supplies $(\Lz/k)^{5+m+n}$ which is precisely right to convert $\tg_m\tp_0^n$ from scaled quantities at $\Lz$ into scaled quantities at $k$. Therefore this shifted operator is respected by the RG at linearised order: 
\eqref{delta-shift+RG} is form invariant under change of scale. Repeating the analysis \eqref{def-physical} and \eqref{physical-dnk}, we thus find that the physical operator also exists and takes the form:
\be 
\label{renormOpshifted}
\lim_{k\to0}\ddp{k}n{\vp+\vp_0} = \ddp{}n{\vp+\vp_0}\,.
\ee

We can connect this observation to the most general form of the physical  potential $V_\p(\vp)$ at the linearised level, when it exists. Indeed for solutions that exist for all $\Lambda\ge0$, we have that 
\be 
\label{general-sol-V}
V(\vp,\Lambda) = \int^\infty_{-\infty}\!\!\!\!\!d\vp_0\, V_\p(\vp_0)\,\ddp\Lambda0{\vp-\vp_0}\,,
\ee
since this clearly satisfies \eqref{flow-V},  whilst from \eqref{renormOpshifted} we see it satisfies the required boundary condition $V(\vp,0)=V_\p(\vp)$. We see that $\ddp\Lambda0{\vp-\vp_0}$ plays the r\^ole of a Green's function, but in \emph{theory space}, giving the form of the potential at any cutoff scale in terms of its \emph{final} functional form. 
%the renormalized potential $V_\p(\vp)$ at $\Lambda=0$. 
By Taylor expanding $\ddp\Lambda0{\vp-\vp_0}$ about $\vp$, we recover the expansion \eqref{expand-V}, 
%written in dimensionful variables:
%\be 
%\label{expand-V}
%V(\vp,\Lambda)= \sum_{n=0}^\infty g_n \,\dd\Lambda{n}\,,
%\ee
but also find a formula for the dimensionful couplings $g_n$ in terms of the physical potential:
\be 
\label{gnVp}
g_n = \frac{(-)^n}{n!}\int^\infty_{-\infty}\!\!\!\!\!d\vp\,\vp^n\, V_\p(\vp)
\ee
(renaming $\vp_0$ as $\vp$).
Actually, substituting the second of \eqref{delta} into \eqref{tg} and using the expression \eqref{On} for the eigenoperator in normal scalar field theory we also have that\footnote{Similarly the couplings \eqref{tg+} in normal field theory can be written as an overlap of the potential with the $\delta_{n}(\tp)$.}
\be 
\tg_n = \frac{(-)^n}{n!}\int^\infty_{-\infty}\!\!\!\!\!d\tp\,\cO{n}(\tp) \tV_\Lambda(\tp)\,,
\ee
which in dimensionful variables gives, using \eqref{physical-OnL},
\be 
\label{gnVL}
g_n = \frac{(-)^n}{n!}\int^\infty_{-\infty}\!\!\!\!\!d\vp\,\pO{n}_{\,\Lambda}(\vp)\, V(\vp,\Lambda)\,.
\ee
Despite appearances, this expression is independent of $\Lambda$ (at the linear level at which we are operating).

Associated to any physical potential $V_\p(\vp)$ is the scale $\Lambda_\p$, which we can now regard as being a dynamical scale characteristic of this particular solution. As before it is defined through the following property of the evolved solution \eqref{general-sol-V}: 
\be 
\label{VpLp}
V(\vp,\Lambda) \in\Lmm \qquad \forall \Lambda>\Lp\,.
\ee
This dynamical scale is the smallest non-negative value satisfying this equation. It can vanish for example if only finitely many $g_n$ are non-vanishing. Since we impose the quantisation condition \eqref{tv-bare}, which then holds for all $\Lambda>\Lz$, a characteristic scale $\Lambda_\p=\infty$  can only be arranged by tuning the $g_n$ in a particular way as the overall UV cutoff is removed.

For $\Lambda<\Lp$, the sum \eqref{expand-V} does not converge. 
%It is tempting to send $\Lambda\to0$ in \eqref{expand-V}, but the resulting expression  ``$V_\p(\vp) = \sum_{n=0}^\infty g_n \,\dd{}n$'' is not well defined. 
However the corresponding expression in conjugate momentum space does make sense. Either from \eqref{dn-Fourier} (with $n=0$, and $\Lz$ replaced with $\Lambda$) and \eqref{general-sol-V}, or directly by Fourier transforming \eqref{flow-V},
\be 
\label{fourier-sol}
V(\vp,\Lambda) = \int^\infty_{-\infty}\!\frac{d\vpi}{2\pi}\, \fV_\p(\vpi)\, {\rm e}^{%\exp\left(
-\frac{\vpi^2}{2}\Omega_\Lambda+i\vpi\vp} \,, %\right)\,,
\ee
where $\fV_\p$ is the Fourier transform of $V_\p$, as is clear by setting $\Lambda=0$. From \eqref{dn-Fourier} and \eqref{expand-V},
\be 
\label{fourier-expansion}
\fV_\p(\vpi) = \sum_{n=0}^\infty g_n (i\vpi)^n\,.
\ee
Since the $g_n$ yield the series \eqref{tv-norm-squared}, which converges for $\Lambda>\Lp$, we see that the above series has an infinite radius of convergence. Therefore $\fV_\p$ is an entire function. Indeed we see that $\Lambda_\p$ characterises the behaviour of the couplings $g_n$  at large $n$, which from \eqref{tv-norm-squared} roughly behave as
\be 
g_n \sim \TRM{\frac{\Lambda_\p^{n+5}}{\sqrt{n!}}}\,.
\ee
The expansion \eqref{fourier-expansion} is the Fourier transform of the formal $\Lambda\to0$ limit of \eqref{expand-V}, \viz  ``$V_\p(\vp) = \sum_{n=0}^\infty g_n \,\dd{}n$''. We see that the expansion of the potential in terms of its eigenoperators is most naturally expressed in conjugate momentum space, through \eqref{fourier-sol} and \eqref{fourier-expansion}.

By \eqref{VpLp} we know that asymptotically we have the leading behaviour for large $\vp$:
\be
\label{asympVpLp}
V(\vp,\Lp)\sim \exp\left(-\frac{a^2\vp^2}{2a^2\Lambda_\p^2}\right) = \exp\left(-\frac{\vp^2}{2\Lambda_\p^2}\right)\,,
\ee
since by assumption the physical potential exists and thus the only allowed reason for exiting $\Lmm$ is the lack of large field convergence in the integral for the norm-squared. Taking the inverse Fourier transform and using \eqref{fourier-sol}, we thus find the $\vpi$ dependence of the physical potential corresponding to this large $\vp$ limit:
\be
\label{fourierVplargephi}
\fV_\p(\vpi) \sim {\rm e}^{-\vpi^2\Lambda_\p^2/4}\,.
\ee
Fourier transforming this gives us the leading asymptotic dependence of the physical potential itself at large $\vp$:
\be 
\label{Vplargephi}
V_\p(\vp)\sim {\rm e}^{-\vp^2/\Lambda_\p^2}\,. %\equiv \exp\left(-\frac{\vp^2}{2\Omega_{\Lp}}\right)\,.
\ee
%where we have used the short-hand $\Omega_\p \equiv \Omega_{\Lambda_\p}$. 
This final result can be confirmed by substituting it into \eqref{general-sol-V}, which recovers \eqref{asympVpLp} but in a way where we clearly rely only on the large field behaviour of $V_\p$. We see therefore that $\Lambda_\p$ is a physical quantity, the \emph{amplitude suppression scale} that characterises the rate of exponential fall-off in the physical potential\footnote{At the linear level, keeping only potential interactions, the Legendre effective potential itself will be universal. In  general such a potential is not universal \cite{Jackiw:1974cv} and instead one must appeal directly to equations of motion \cite{Nielsen:1975fs}.} at large $\vp$. Our reason for including the non-universal factor $a$ in \eqref{VpLp} (and similar earlier equations) is finally apparent: it is so that $\Lambda_\p$ in this case is indeed universally related to a physical quantity. From here on we take \eqref{Vplargephi} as the primary definition $\Lambda_\p$,  whenever the physical potential exists. In sec. \ref{sec:compact} we will see another physical consequence of this scale.
If we restore $\hbar$, it sits in front of $\Omega_{\Lp}=\Lambda_\p^2/2$. Therefore \eqref{Vplargephi} establishes that even outside $\Lmm$ the potential, and in particular the physical potential, remains non-perturbatively quantum.

Since \eqref{fourier-sol} is the general solution of \eqref{flow-V}, it gives the RG flow starting from any bare potential $V(\vp,\Lz)$, except
 of course that $\fV_\p$ is no longer the Fourier transform of the physical potential if the flow ends prematurely. Rewriting the solution in terms of the Fourier transform of the bare potential, we have
\be 
\label{fourier-sol-fromBare}
V(\vp,\Lambda) = \int^\infty_{-\infty}\!\frac{d\vpi}{2\pi}\, \fV(\vpi,\Lz)\, %{\rm e}^{%
\exp\left(
\frac{\vpi^2}{4a^2}(\Lzp2-\Lambda^2)+i\vpi\vp %} \,. %
\right)\,.
\ee
From this expression we see clearly why a generic choice of bare potential leads to the flow ending in a singularity: for sufficiently small $\Lambda$ the integrand diverges at large $\vpi$. If the integral fails to converge first at $\Lambda=\Lp$, then precisely at this point the typical result will be a distributional $V(\vp,\Lp)$.

\subsection{Examples at first order in the couplings}
\label{sec:examples}

\begin{figure}[ht]
\begin{center}
$
\begin{array}{cc}
\includegraphics[width=0.45\textwidth]{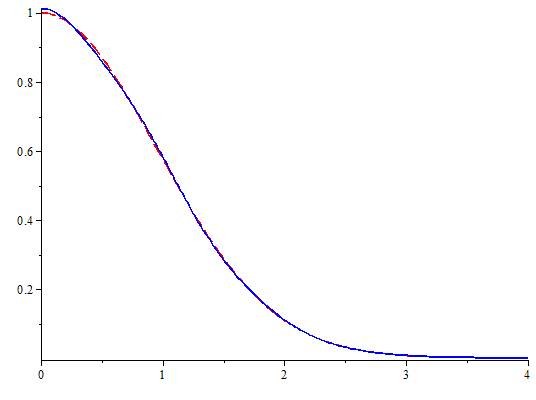} &
\includegraphics[width=0.45\textwidth]{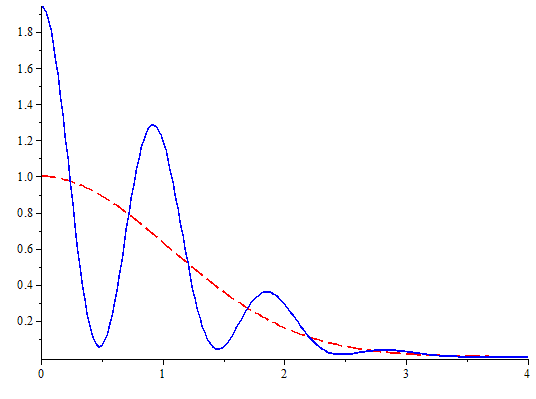} \\[-0.3cm]
a\tp & a\tp
\end{array}
$
\end{center}
\caption{Plotted in dashed red is the exact potential \eqref{exampleVL} normalized to $V(0,\Lambda)=1$, and in solid blue its finite sum up to and including $g_{20}$. The left panel is the situation when $\tilde{\Lambda}_\p=0.9$, \ie just inside $\Lmm$, while the right panel is the situation having just exited, with $\tilde{\Lambda}_\p=1.1$.}
\label{fig:completeness}
\end{figure}

The simplest example nevertheless illustrates and confirms the general behaviour derived above. We need an entire function for $\fV_\p$. We take just \eqref{fourierVplargephi} with coefficient $\Lambda_\p^5\sqrt{\pi}$, consistent with dimensions. Then
\be 
\label{exampleVp}
V_\p(\vp) = \Lambda_\p^4\, {\rm e}^{-{\vp^2}/{\Lambda_\p^2}} \,,
\ee
while from \eqref{fourier-expansion}, the odd-$n$ couplings vanish and the even-$n$ ones are given by
\be 
\label{exampleg2m}
g_{2m} =\frac{\sqrt{\pi}}{m!4^m} \Lambda_\p^{5+2m}\,. %\, \left(\frac{\Lp}{2a}\right)^{2m}\,.
\ee
One can confirm that these couplings are reproduced by \eqref{gnVp}, or \eqref{gnVL} using the formula below.
Performing the integral in \eqref{fourier-sol} gives the evolved potential:
\be 
\label{exampleVL}
V(\vp,\Lambda) = \frac{a\Lambda_\p^5}{\sqrt{\Lambda^2+a^2\Lambda^2_\p}}\, \exp\left(-\frac{a^2\vp^2}{\Lambda^2+a^2\Lambda^2_\p}\right)\,.
\ee
We see explicitly that $V(\vp,\Lambda)\in\Lmm$ only for $\Lambda>\Lp$, exiting at $\Lp$ through failure of the integral to converge at large $\vp$. Computing the norm-squared integral gives 
\be 
\label{example-norm-squared}
\frac{\sqrt{\pi}\,\tilde{\Lambda}^{10}_\p}{a^9\sqrt{1-\tilde{\Lambda}^4_\p}}\,,
\ee
where $\tilde{\Lambda}_\p = \Lp/\Lambda$, which indeed can be expressed as the series in \eqref{tv-norm-squared} when $\Lambda>\Lp$.  The Hilbert space property, in particular \eqref{completeness-proof-}, is illustrated in fig. \ref{fig:completeness}, by comparing the exact result \eqref{exampleVL} to the finite sum, namely \eqref{expand-V} with the upper limit replaced by $N=20$.

We can take the bare potential to be \eqref{exampleVL} for any $\Lambda=\Lz>\Lp$. Qualitatively, the property it has that allows it to survive all the way down to $\Lambda=0$, is that it is at least as spread out as the eigenoperators themselves (although if it is more spread out, then it exits $\Lmm$ through failure of the integral to converge at large $\vp$ as we have seen). In particular therefore for a physical potential to exist, the bare potential $\tV_\Lz(\tp)\in\Lmm$ must decay for large $\tp$ as 
%\be 
%\label{bare-decay}
$\exp(-a_0^2\,\tp^2)$, 
%\,, \ee
where $1/2<a^2_0/a^2\le 1$, but also there can be no smaller-scale features in the bare potential.

On the contrary, if we take a bare potential with finer features than the eigenoperators, taking for example the more compact ($\Lz>\Lp$):
\be 
\label{exampleVL0}
V(\vp,\Lambda_0) = \frac{a\Lambda_\p^5}{\sqrt{\Lambda_0^2-a^2\Lambda^2_\p}}\, \exp\left(-\frac{a^2\vp^2}{\Lambda_0^2-a^2\Lambda_\p^2}\right)\,,
\ee
then the flow fails before reaching $\Lambda=0$. By comparing to \eqref{exampleVL}, we see that for this example $V(\vp,\Lambda)$ is just given by the above expression with $\Lz$ replaced by $\Lambda$. The couplings $g_{2m}$ are then those of \eqref{exampleg2m} but with a $(-)^m$ factor on the right hand side, and the norm-squared integral is the same as \eqref{example-norm-squared}. However this time the exit from $\Lmm$ is due to the fact that as $\Lambda$ approaches $\Lp$, the width of the exponential vanishes, indeed
\be 
\label{sticky-end}
\lim_{\Lambda\to\Lp^+} V(\vp,\Lambda) = \Lambda^5_\p {\sqrt{\pi}}\, \delta(\vp)\,.
\ee
Attempting to flow below this point by analytic continuation gives a complex answer in general, in this case pure imaginary:
\be 
\label{imaginary-V}
V(\vp,\Lambda) = i \frac{a\Lambda_\p^5}{\sqrt{a^2\Lambda^2_\p-\Lambda^2}}\, \exp\left(\frac{a^2\vp^2}{a^2\Lambda^2_\p-\Lambda^2}\right)\,,\qquad\Lambda<\Lp\,.
\ee

For completeness, let us mention that by using \eqref{general-sol-V} and an appropriate choice of $V_\p$, one can generate flows $V(\vp,\Lambda)$ that exist for all $\Lambda\ge0$ but which never enter $\Lmm$. For example choose 
\be 
\label{bad1}
V_\p(\vp) = \frac1{\Lambda_\p^2+\vp^2}\qquad\implies\qquad
%\ee
%we get the Fourier transform
%\be 
\fV_\p(\vpi) = \frac{\pi}{\Lambda_\p}\,{\rm e}^{-\Lambda_\p|\vpi|}\,.
\ee
Since the latter has no Taylor expansion, the couplings do not exist, \cf \eqref{fourier-expansion}. By \eqref{general-sol-V} or \eqref{fourier-sol}, 
\be 
V(\vp,\Lambda) = \frac{a\sqrt{\pi}}{\Lambda\Lambda_\p} \,{\rm Re} \left\{ {\rm e}^{(\tilde{\Lambda}_\p+ia\tp)^2} {\rm Erfc}(\tilde{\Lambda}_\p+ia\tp)\right\}\,, %= \frac1{\Lambda_\p^2+\vp^2} +O\left(\frac1{\vp^4}\right)\ \notin\Lmm\,,
\ee
%the second equality holding for $|\vp|\gg\Lambda_\p$. 
whose large $\vp$ behaviour is the same as at $\Lambda=0$, \ie \eqref{bad1}. On the other hand, choose
\be 
\label{bad2}
\fV_\p(\vpi) = \frac1{1+\Lambda_\p^2\vpi^2}\qquad\implies\qquad
V_\p(\vp) = \frac{\pi}{\Lambda_\p}\,{\rm e}^{-|\vp|/\Lambda_\p}\,.
\ee
In this case, the couplings exist ($g_n = \Lambda_\p^n\, \delta_{n={\rm even}}$) but clearly from \eqref{tv-norm-squared}, $V(\vp,\Lambda)$ is never in $\Lmm$. Indeed from \eqref{general-sol-V} one finds its large $\vp$ behaviour is again unchanged from what it was at $\Lambda=0$, namely \eqref{bad2}. In both cases $V$ is never in $\Lmm$ because its large $\vp$ decay is too weak for all $\Lambda$. The difficulty is making physical sense out of these behaviours. In the latter case, Green's functions and $S$ matrix elements do not exist because $V_\p$ is not differentiable at $\vp=0$. In both cases, there is no well defined way to isolate relevant and irrelevant parts and thus to define what one means by the continuum limit.

\subsection{Derivative eigenoperators}
\label{sec:derivative-ops}

Now we derive the form of the general eigenoperator, with spacetime derivative interactions. It will be sufficient to consider adding kinetic term interactions 
%(with the appropriate sign) 
to \eqref{linearised-v}, to see the general pattern. Thus we set:
\be 
\label{linearised-k}
\delta S^\Lambda = -\epsilon\! \int\!\!d^4x\,\left\{ V\!\left(\ph(x),\Lambda\right)+ \frac12\left(\partial_\mu\vp\right)^2K\!\left(\ph(x),\Lambda\right)\right\} \,.
\ee
Recall that the linearised flow is the same whether we consider this to be part of the Wilsonian or Legendre effective action.  Note the overall sign. In view of the negative sign kinetic term, this is the natural sign for the interactions, \ie assuming $K>0$. Up until now the overall sign of the potential term in the action, has not mattered,\footnote{The equations in the previous subsection are blind to this sign.} however classical stability \TRM{would now require} that the potential \TRM{is} bounded above.\footnote{Without \TRM{this}, the consequent classical instability also leads \TRM{inevitably} to a pole in \eqref{Gamma.flow-}.}  \TRM{Changing its sign as in \eqref{linearised-k} then returns it to being bounded below.} Working in scaled variables \eqref{scale}, $K=\tK$, the eigenoperators are defined by the $K$ component:
\be 
\tK(\tp,t) = \left(\frac{\mu}{\Lambda}\right)^\lambda \tK(\tp)
\ee
and $V$ component \eqref{lambda-v}.  We thus find the simultaneous equations:
\bea 
\label{eigenK}
-\lambda\, \tK(\tp) - \tp\, \tK' &=& \frac{\tK''}{2a^2}\,,\\
\label{eigenK-V}
-\lambda\, \tilde{V}(\tp) -\tilde{\ph}\, \tilde{V}' + 4\, \tilde{V} &=& \frac{\tilde{V}''}{2{a}^2}+2b\tK\,,
\eea
where we have set
\be 
b = \int\!\frac{d^4\tilde{p}}{(2\pi)^4}\, C(\tilde{p}^2)\,.
\ee
Of course we still have the solutions $\tV(\tp) = \delta_n(\tp)$, $\tK(\tp)=0$. We also clearly have solutions $\tV= b \tK/2$. By comparing to \eqref{eigen-}, we see that these $O(\partial^2)$ eigenoperators thus take the form:
\be 
-\frac12\,  \delta_n(\tp)\left [\left(\tilde{\partial}_\mu\tp\right)^2 + b\, \right]\,,\qquad \lambda= 1+n\,,
\ee
implying that these operators have dimension $3-n$. Clearly the $\tK$ and $\tV$ parts are  in $\Lmm$. We can extend the definition of $\Lmm$ by stripping off the purely space-time derivative parts in this way. All the other (polynomial and Kummer function) solutions to \eqref{eigenK} and \eqref{eigenK-V} lie outside $\Lmm$ and thus are excluded from the bare action. Importantly note that the kinetic term $\left(\tilde{\partial}_\mu\tp\right)^2$ is not itself an eigenoperator, since a constant is not integrable under ${\rm e}^{+a^2\tp^2}$.

Equivalently we can define $\Lmm$ to be the space of interactions that are integrable under ${\rm e}^{+a^2\tp^2_0}$, where we shift the field by a spacetime independent constant, $\tp(\tilde{x})\mapsto\tp(\tilde{x})+\tp_0$. So far we have been assuming that the interaction is localised, \ie all fields in the interaction have the same spacetime argument $x$. This latter definition of $\Lmm$ allows us to extend it to non-local interactions, although such an interactions can only then be expanded in terms of the eigenoperators if they are quasi-local \ie possess a space-time derivative expansion.

Like the potential operators $\delta_n$, these $O(\partial^2)$ operators are all relevant, and thus all associated with renormalized couplings in the continuum limit (in this case $\tg_n = g_n/\Lambda^{1+n}$). Since $b>0$, the associated potential contribution has naturally the right sign for classical stability. As might have been expected, given that these eigenoperators are defined at a Gaussian fixed point, their scaling dimension equals the sum of the dimensions of the components:
\be 
\label{dimension-rule-K}
3-n = [\left(\partial_\mu\vp\right)^2]+\left[\delta_n\right]\,,
\ee
where the scaling dimension of the first term is also its engineering dimension, and the second is given by \eqref{delta-dim}. The dimensionful operators are given by multiplying by $\Lambda^{3-n}$ and thus take the form:
\be 
\label{K-physical}
-\frac12\,  \dd\Lambda{n}\left [\left({\partial}_\mu\vp\right)^2 + b\Lambda^4\, \right]\,,
\ee
and consequently, taking the IR limit $\Lambda\to0$, the physical operators are:
\be 
\label{K-renormalised}
-\frac12\,  \dd{}{n}\left({\partial}_\mu\vp\right)^2 \,.
\ee

It is straightforward to see how this generalises to arbitrary derivative interactions. We add to the effective Lagrangrian a term
\be 
\epsilon L\!\left(\ph,\Lambda\right) \mon_p(\partial,\partial\ph)\,,
\ee
%%https://en.wikipedia.org/wiki/Algebraic_fraction
where $\mon_p$ is some Lorentz invariant monomial with $2p$ space-time derivatives, of definite engineering dimension $d_p$, and where each instance of $\vp$ appears differentiated at least once. Tadpole corrections will generate subleading terms $\mon_{0\le\, p'<p}$ of lower dimension $d_{p'}$, which thus must also be added, together with their coefficient functions. For the eigen-functions, the top function, $\tilde{L}(\tp)$, satisfies the same equation as \eqref{eigen-} except that by scaling as in \eqref{scale}, the dimension $4$ is replaced by $4-d_p$. We thus find that the interactions in $\Lmm$ are again formed by setting $\tilde{L}(\tp)\propto \delta_n(\tp)$, where they form a basis for such $\mon_p$ interactions. Similarly to \eqref{dimension-rule-K} their dimensions are thus $d_p-1-n$, while the dimension of the associated coupling is $5+n-d_p$. Thus again infinitely many of this tower of higher derivative operators are relevant. However for $d_p\ge 5$, the $n=d_p-5$ operator is marginal. And once $d_p\ge6$, those $n< d_p-5$ operators are irrelevant, and thus in the continuum limit have couplings that are determined by the relevant ones. The coefficient functions for the subleading terms will satisfy equations somewhat similar to \eqref{eigenK-V}, for which we want the special solution which will be tied to $\delta_n(\tp)$. Since their dimension $d_{p'}<d_p$, they will appear in the dimensionful eigenoperators with positive powers of $\Lambda$ like in \eqref{K-physical}. Finally the physical operators will simply be %\footnote{Without loss of generality, we have set the proportionality constant to unity.}
\be 
\label{derivative-operator-basis-renormalised}
\dd{}n\,\mon_p(\partial,\partial\ph)\,.
\ee
We see that the novel physical properties, namely non-perturbative in $\hbar$, evanescence and IR suppression, are also true of all the derivative interactions. %On the other hand this all follows from the $\delta_n$: a
Apart from the role of the polynomial basis \eqref{On} now being played by $\delta_n(\tp)$, this structure closely mimics that of scalar field theory with positive kinetic term. Similarly therefore, we anticipate that a more convenient basis for the Hilbert space of interactions,
%(for $\Lm\pm$) 
is to use the top term and discard the subleading corrections:\footnote{although we are discarding only the $\mon_{p'<p}$ terms, not the crucial tadpole corrections to $\dd{}n$. Of course the maximal subset of $\sigma_p$ should be chosen so that \eqref{derivative-operator-basis-physical} are independent under integration by parts.}
\be 
\label{derivative-operator-basis-physical}
\dd\Lambda{n}\,\mon_p(\partial,\partial\ph)\,,
\ee
and with a slight abuse of terminology, classify these as relevant, marginal, or irrelevant.  
Thus for example we recognise that $\dd\Lambda0\,(\Box\vp)^2$ is an irrelevant operator, $\dd\Lambda1\,(\Box\vp)^2$ is marginal, and all the  $\dd\Lambda{n>1}\,(\Box\vp)^2$ are relevant.

\section{Perturbation theory}
\label{sec:perturbation-th}

We have seen that already at the linear level, the structure is non-perturbative in $\hbar$, but nevertheless calculable. This is also true for corrections which can be developed as a perturbation theory in the couplings $g_n$, while staying non-perturbative in $\hbar$. That this can be done consistently, rests upon the fact that, term by term, the corrections remain in $\Lmm$.
%We will not try to give a comprehensive account of this here, but instead try to draw out some of the salient features. % before applying these ideas to quantising gravity.}
%Beyond the linear order, when we consider perturbation theory, we will obtain
Indeed, in these terms we will find differentials of the eigenoperators, which by \eqref{d-delta} trivially remain in $\Lmm$. As we will see in sec. \ref{sec:QG}, when applied to quantum gravity we can expect to obtain terms with $\delta_m(\tp)$ times a positive integer power of $\tp$. 
%The power is either one or two, but clearly for any power, t
This is again in $\Lmm$. In fact from \eqref{dn-Fourier} it is straightforward to derive
\be 
\label{dn-phi}
\vp\, \dd\Lambda{n} = -n\, \dd{\ \Lambda}{n-1}\,-\,\Omega_\Lambda\, \dd{\ \Lambda}{n+1}
\ee
(which from \eqref{delta} is just the Hermite polynomial recurrence relation in disguise). 

Finally, we will also
obtain products of the eigenoperators. Clearly such products are again in $\Lmm$, and thus, if quasi-local, we can expand them back into the eigenbasis. We are thus faced generically with
\be 
\label{prod}
\delta_m(\tp)\,\delta_n(\tp) =\sum_{j=0}^\infty\cc{mn}j\, \delta_j(\tp)
\ee
(where the fields are all at the same spacetime point).
From \eqref{tg} and a Hermite linearization formula \cite{Gradshteyn1980},  the expansion coefficients are:
\be 
%\tg_j \equiv 
\cc{mn}j=\frac{2^{s-j}a^{2s-2j}}{2\pi^2j!} \Gamma(s-j)\Gamma(s-m)\Gamma(s-n)\, \delta_{j+m+n\,=\,{\rm even}} \,,\quad {\rm where}\quad 2s = j+m+n+1\,.
\ee
However, using Stirling's formula for large $j$, we find
\be 
j! \left(\cc{mn}j\right)^2 \sim \frac{a^{2(m+n+1)}}{\sqrt{2\pi^3}} \frac{\ j^{m+n-\half}}{(4a^2)^j}\,,
\ee
therefore we see that this is a case where \eqref{tv-norm-squared} has a finite radius of convergence. Assuming for the moment that \eqref{prod} appears in the bare action, thus with %small enough 
coupling $\tg_{mn}=g_{mn}/\Lzp{6+m+n}$, and we evolve the product itself at the linearised level (this is not exactly how it arises, but this discussion will be useful shortly),
it leaves $\Lmm$ for $\Lambda\le \Lp$ where
\be 
\label{prodLp}
\Lp =\Lz/\sqrt{2}\,. 
\ee
To see this  we note that the corresponding dimensionful coefficients are:
\be 
\label{c-dim}
c^j_{mn} := \cc{mn}j\,\Lzp{j-m-n-1}\,,
\ee
and then we use \eqref{tv-norm-squared} to compute the norm-squared at scale $\Lambda$.
Having defined the dimensionful coefficients by \eqref{c-dim}, the dimensionless expansion evolves self-similarly, in particular $\tc{mn}j = c^j_{mn}/k^{j-m-n-1}$,  this fact being guaranteed for the couplings by dimensional analysis. However the relation \eqref{prod} is not respected by the RG already at linearised level: the evolved expansion
\be 
\label{prod-evolved}
\left[\delta_m(\tp)\,\delta_n(\tp)\right]^\Lz_k := \sum_{j=0}^\infty\tc{mn}j\, \delta_j(\tp)\,,
\ee
is only equal to $\delta_m(\tp)\,\delta_n(\tp)$ at the original scale $k=\Lz$.

Since the $\cc{mn}j$ are pure numbers, we see that the relevant couplings $g_{mn}\, c^j_{mn}$ are large (for large enough $j$), as set by the bare cutoff scale $\Lz$. Since (at finite scales) the relevant couplings must be finite in the continuum limit, we see that we would need to compensate by adjusting the bare values of $g_j$, in other words they would need renormalization. In fact the single term $g_{mn}\delta_m(\tp)\,\delta_n(\tp)$ in the bare potential is anyway unacceptable at the linearised level, because such a potential is more compact than the eigenoperators. Thus the flow in fact ends at \eqref{prodLp} with a distributional effective potential. Indeed the bare potential can be rewritten in this case as 
\be 
P\left(\partial_\vp\right)\,\left(\dd\Lz0\right)^2\,,
\ee
where the first term is a rank $m+n$ polynomial of $\vp$ derivatives. The second term is proportional to
\eqref{exampleVL0}, with $\Lp$ again given by \eqref{prodLp}, and thus  the whole combination evolves to this constant of proportionality times $P\left(\partial_\vp\right)$ acting on \eqref{sticky-end}.

%Armed with these properties, we can now develop the perturbation theory. 
%We begin to develop the perturbation theory. 
Now we demonstrate how perturbation theory can be developed. 
Since we need results that are non-perturbative in $\hbar$, we must in effect sum over all Feynman diagrams to infinite order. What promises to keep this manageable is that we can nevertheless expand perturbatively in the couplings. To get insight we first proceed this way, working directly from the functional integral. Then we will turn to solving the flow equations, which provides a more elegant and more powerful approach for our purposes.

\subsection{Second order in the couplings by summing Feynman diagrams}
\label{sec:pert2-textbook}

%% see e.g.(206.9), (197.10) etc

\begin{figure}[ht]
\centering
\includegraphics[scale=0.4]{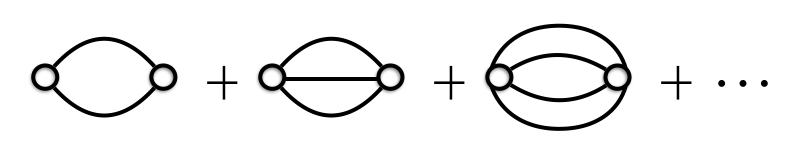}
%\vskip-30pt
\caption{Feynman diagrams at second order in the coupling but all orders in $\hbar$.}
\label{fig:melons}
\end{figure}

At second order in the couplings, the 1PI contribution will be computed from all such Feynman diagrams involving two bare operators at spacetime points $x_1$ and $x_2$, each taking the form of \eqref{derivative-operator-basis-physical} with $\Lambda=\Lz$. If for illustrative purposes we keep all and only the non-derivative operators, then this can be written as the $\vp$ dependent 1PI part of the functional integral
\be 
\label{V2nd}
%\frac{1}{\mathcal{Z}_0}
\frac12\int\!\! \mathcal{D}\vp_q\ {\rm e}^{\frac{1}{2}\varphi_q\cdot \left(\Delta_{k}^\Lz\right)^{\!-1}\!\!\!\!\!\cdot \varphi_q}\,  \int_{x_1} V\!\left(\vp_q(x_1)\!+\!\vp(x_1),\Lz\right)\,\int_{x_2} V\!\left(\vp_q(x_2)\!+\!\vp(x_2),\Lz\right)\,.
\ee
The  exponential of the fluctuation field $\vp_q(x)$ has the wrong sign for promoting convergence.  As mentioned at the beginning of sec. \ref{sec:minus}, at first sight this makes no sense and, as is clear from \eqref{sum-ruled-Delta}, the exponential divergence gets dramatically worse as $k\to\Lz$, rather than suppressing the integral. However this latter divergence belongs only to the field independent part and we are not interested in that. 
%we see again the problem which will lead generically to the flow failing at some $0<k<\Lz$.
By using \eqref{fourier-sol} at $\Lambda=\Lz$, the dependence on the fluctuation field from the interactions can be isolated through ${\rm e}^{iJ\cdot\vp_q}$, where 
\be 
J(z) = i\sum_{j=1,2}\vpi_j\,\delta(x_j-z)\,,
\ee
and $\vpi_j$ is the corresponding conjugate momentum. Performing  the now-Gaussian functional integral gives
\be 
\label{V2ndResult}
\frac12\int_{x_1,x_2}\int \frac{d\vpi_1d\vpi_2}{(2\pi)^2}\, \fV_\p(\vpi_1,\Lz)\fV_\p(\vpi_2,\Lz)\, {\rm e}^{-\frac12\vpi_iM_{ij}\vpi_j+i\vpi_i\vp(x_i)} \Big|_{\rm 1PI}\,.
\ee
Anticipating that the dimensionful couplings $g_n$ will now run with scale, we set them to their bare values $g_n(\Lz)$, or equivalently through \eqref{fourier-expansion}, set $\fV_\p$ to its bare value. We have also introduced the $O(\hbar)$ $2\!\times\!2$ matrix 
\be 
M = \begin{pmatrix}
\Omega_k & -\Delta^{\!\Lz}_k(x_1,x_2)\\
-\Delta^{\!\Lz}_k(x_1,x_2) & \Omega_k
\end{pmatrix}\,.
\ee
The $\Omega_k$ entries arise in the same way as in \eqref{tadpoles-evolution}, and thus re-sum the tadpole graphs in fig. \ref{fig:tadpoles}, turning the constituent bare eigenoperators into renormalized ones. Expanding perturbatively in $\Delta^{\!\Lz}_k(x_1,x_2)$ generates the graphs in fig. \ref{fig:melons} that connect the two renormalized eigenoperators. Finally, the restriction to 1PI means that one should subtract the terms zeroth and first-order in $\Delta^{\!\Lz}_k(x_1,x_2)$.

If individual eigenoperator contributions were representative of the whole, for example if only a finite number of couplings were non-vanishing, we see via \eqref{fourier-expansion} that the $\vpi$ integral in \eqref{V2ndResult} would diverge as soon as $M$ is no longer positive definite. Since $\Delta^{\!\Lz}_k(x_1,x_2)$ is a decreasing function of $|x_1-x_2|$,\footnote{This is \eg clear from the fact that $\Delta^{\!\Lz}(r)-\Delta^{\!\Lz}(r') > \Delta^k(r)-\Delta^k(r')$ for $r=|x_1-x_2|<r'=|x'_1-x'_2|$.} this happens first at coincident points where
\be 
\label{coincidentDelta}
\Delta^{\!\Lz}_k(x_1,x_1) = \Omega_\Lz - \Omega_k = \frac{\Lz^2-k^2}{2a^2}\,,
\ee
meaning that $k$ could not be lowered below $\Lz/\sqrt{2}$, as in \eqref{prodLp}. We recognise that the flow has broken down for the reasons given in the previous subsection. 

But operator mixing will switch on all couplings, which furthermore will run with scale. Their bare values will be weighted by the appropriate power of $\Lz$ as set by dimensions (but such that the couplings nevertheless behave correctly so as to access the Gaussian continuum limit). At the bare level, for large $\vpi$, we therefore expect something like
\be 
\fV_p(\vpi,\Lz) \sim {\rm e}^{-\vpi^2\Lzp2/4c^2_0}\,,
\ee
for some bare coefficient $c_0(\Lz)>0$ (compare \eqref{fourierVplargephi}). Then providing $c_0<a$,  the same arguments as in \eqref{coincidentDelta} show that \eqref{V2ndResult} would be well defined for all $k\ge0$. However, as well as resorting to guesswork, we are also ignoring the  contributions from the (marginally) relevant derivative operators \eqref{derivative-operator-basis-physical}, all of which will also contribute.

\subsection{Second order in the couplings by solving the flow equation}
\label{sec:pert2-flow}

This complexity is much better handled by solving the flow equations directly. The simplest description arises from taking the 1PI part $\Gamma_\Lambda:=\Gamma_\Lambda^\infty$ of the Wilsonian effective action $S^\Lambda$  \cite{Morris:1993,Morris:1998,Morris:2015oca} since this will give us direct access to the $\beta$ functions induced by quantum corrections, and involves only the one scale, $\Lambda$. At the same time this solves for the IR cutoff Legendre effective action directly in the continuum limit. Writing $\Gamma^{(n)}$ to be the part $n^{\rm th}$ order in the couplings, and expanding the right hand side of \eqref{Gamma.flow-} to second order in the couplings, we have $\Gamma_\Lambda=\Gamma^{(1)}+\Gamma^{(2)}$, where\footnote{and from \eqref{DeltaUV} and \eqref{sum-rule}, $\Delta_\Lambda(p) =\Delta^\infty_\Lambda(p)= [1-C^\Lambda(p)]/p^2$.}
\be 
\label{flow-2nd}
\dot{\Gamma}^{(1)}[\varphi]+\dot{\Gamma}^{(2)}[\varphi] %-\sum^\infty_{n=0} \dot{g}_n\, \int_x \dd\Lambda{n} =
= -\frac{1}{2}\,\text{tr}\left[\dot{\Delta}_\Lambda\cdot \frac{\delta^{2}\Gamma^{(1)} }{\delta\ph\delta\ph}\right] 
-\frac{1}{2}\,\text{tr}\left[\dot{\Delta}_\Lambda\cdot \frac{\delta^{2}\Gamma^{(2)} }{\delta\ph\delta\ph}\right] 
-\frac12\,\text{tr}\left[\dot{\Delta}_\Lambda\cdot\frac{\delta^{2}\Gamma^{(1)} }{\delta\ph\delta\ph}\cdot\Delta_\Lambda\cdot\frac{\delta^{2}\Gamma^{(1)}}{\delta\ph\delta\ph} \right]
\ee
As we have already emphasised, we need to work non-perturbatively in the loop expansion. It is therefore important to recall that the flow equations \eqref{pol-} and \eqref{Gamma.flow-} are indeed non-perturbative, in fact exact, RG equations. Written in the form \eqref{flow-2nd} the flow equation is now second order in the couplings,  but it is still exact in $\hbar$. If we were to solve \eqref{flow-2nd} by iteration, we would reproduce the Feynman diagrams just considered, in particular the last term gives those in fig. \ref{fig:melons}.

Now we again concentrate on the potential. We have seen that at first order we have the solution
\be 
\label{linearised}
\Gamma_\Lambda[\vp] = \Gamma^{(1)}=
-\int_x V\!\left(\vp(x),\Lambda\right)\,, 
%\qquad\text{where}\qquad V(\vp)  = \sum^\infty_{n=0} g_n\, \dd\Lambda{n}\,,
\ee
where $V$ is given by \eqref{fourier-sol}, for some $\Lambda$-independent $\fV_\p$, which when expanded as in \eqref{fourier-expansion} gives thus $\Lambda$-independent $g_n$. If the flow survives down to $\Lambda=0$, then $\fV_\p$ is the Fourier transform of the resulting physical potential $V_\p$. When $V(\vp,\Lambda)\in\Lmm$, we can instead expand it directly, as in \eqref{expand-V}. Beyond linearised order, we need to define  the couplings by an appropriate renormalization condition. Since the IR cutoff ensures that $\Gamma_\Lambda$ has a spacetime derivative expansion, we choose to define the $g_n$ to be the Taylor expansion coefficients of the corresponding $\fV_\p$, which thus now runs:
\be 
\label{fourier-expansion-running}
\fV_\p(\vpi,\Lambda) = \sum_{n=0}^\infty g_n\!(\Lambda) \left(i\vpi\right)^n\,.
\ee
%in the $O(\partial^0)$ part written in conjugate momentum space as in \eqref{fourier-expansion}. 
While $V\in\Lmm$, this is equivalent to requiring that $g_n(\Lambda)$ is the coefficient of the operator $\dd\Lambda{n}$.

By the renormalization conditions, $\Gamma^{(2)}$ has no interaction potential.
Thus the only piece that contributes to the running of the potential is the $O(\partial^0)$ part of the final term which evaluates to $c \int_x \left(\partial^2_\vp V\right)^2$, 
%where prime means $\partial/\partial\vp$.
%which thus induces the couplings $g_n$ to run. 
where $c$ is a universal term, the one-loop diagram:
\be 
c = -\frac12\int\!\frac{d^4p}{(2\pi)^4}\, \Delta_\Lambda(p)\dot{\Delta}_\Lambda(p) = -\frac{1}{32\pi^2} \int^\infty_0\!\!\!\!\!\! dp\,\, \frac{\partial}{\partial p} C^2_\Lambda = -\frac{1}{32\pi^2} \,.
\ee
By \eqref{physical-dnk} and \eqref{expand-V}, %field derivatives just increase the index $n$. 
while $V\in\Lmm$ we have
\be 
\label{Vpp}
\partial^2_\vp V(\vp,\Lambda)  = \sum^\infty_{n=0} g_n\, \dd\Lambda{n+2}\,.
\ee
Converting to scaled operators using \eqref{def-physical}, using the product formula \eqref{prod}, and then converting back we thus find
\be 
\label{beta-physical}
%\Lambda \frac{\partial}{\partial\Lambda} 
\dot{g}_j = \frac{\Lambda^{j-5}}{32\pi^2}\sum_{m,n=0}^\infty \frac{\cc{m+2,n+2}j }{\Lambda^{m+n}} g_mg_n\,,
\ee
or in autonomous form:
\be 
\label{beta-scaled}
\Lambda \frac{\partial}{\partial\Lambda} \tg_j = -(5+j)\tg_j-\frac{1}{32\pi^2}\sum_{m,n=0}^\infty {\cc{m+2,n+2}j } \tg_m\tg_n\,.
\ee
Relying on the existence of flows in the reverse direction, we can now solve these equations for $\Lambda>\mu$ for any given choices of `initial' couplings $g_j(\mu)$. Indeed \TRM{it} is straightforward to solve \eqref{beta-physical} as a perturbative series in powers of $g_j(\mu)$:
\be 
g_j(\Lambda) = g_j(\mu) + \frac{1}{32\pi^2}\sum_{m,n=0}^\infty \frac{\cc{m+2,n+2}j }{m+n+5-j}g_m(\mu)g_n(\mu) \left(\Lambda^{j-m-n-5}-\mu^{j-m-n-5}\right) + O\left(g^3(\mu)\right)\,.
\ee
Note that since $\tg_j(\Lambda) =g_j(\Lambda)/\Lambda^{j+5}$, order by order in the perturbation theory all these solutions emanate  from the Gaussian fixed point in the $\Lambda\to\infty$ limit as required.

We have  only kept track of the $O(\partial^0)$ parts.\footnote{We cannot therefore directly compare this to the calculation in sec. \ref{sec:pert2-textbook}, where the induced higher derivative contributions are implicitly included at scales $k<\Lz$, through $\Delta^\Lz_k(x_1,x_2)$.} The last term in \eqref{flow-2nd} provides a spacetime derivative expansion to all orders. Expanding these into the basis \eqref{derivative-operator-basis-physical}, it will contribute to the $\beta$ functions for all the other relevant couplings. A continuum limit can therefore be achieved only by working simultaneously with all the relevant couplings, as expected on general grounds. Defining their renormalization conditions in a similar way, will mean that $\Gamma^{(2)}$ contains no relevant operators. Its only purpose is to solve for the couplings of the irrelevant operators which, in the continuum limit, are determined by the irrelevant operator parts extracted from the last term. Of course once we recognise that all the other relevant couplings must be switched on, the second-order $\beta$ functions above will receive contributions from them as well. 

We note that the arbitrarily negative powers of $\Lambda$ that appear in \eqref{beta-physical} prevent a smooth $\Lambda\to0$ limit existing, unless all the couplings $g_n$ vanish in this limit. To show this we assume a $\Lambda\to0$  limit does exist for which $V(\vp,0)\ne0$ and show that $\partial_\Lambda V(\vp,\Lambda)$ must then diverge in this limit. First note that outside $\Lmm$, we would get the same formula by using \eqref{fourier-expansion-running} and \eqref{fourier-sol} and Fourier transforming the final $c \int_x \left(\partial^2_\vp V\right)^2$ term. 
In fact having isolated the $O(\partial^0)$ part, this last term is the only term that survives the $\Lambda\to0$ limit on the right hand side of \eqref{flow-2nd}, and is non-vanishing if the couplings are non-vanishing in this limit. This implies that $\Lambda\partial_\Lambda V(\vp,\Lambda)$ has a finite limit, which in turn implies that $\partial_\Lambda V(\vp,\Lambda)$ itself must diverge in the $\Lambda\to0$ limit. 

However, as we will address in sec. \ref{sec:infinite},  these couplings generate a mass $m$, which must  then be handled non-perturbatively.
Then it is no longer true that the evolution of the couplings $g_j$ are tied to the scale $\Lambda$ and we can expect that they generically freeze out at values set by the scale $m$, as $\Lambda\to0$. We similarly expect finite size effects (see sec. \ref{sec:compact})  to provide a freeze-out scale $1/L$ on a sufficiently homogeneous manifold.

\subsection{Higher orders and infinite order}
\label{sec:infinite}

\begin{figure}[ht]
\centering
\includegraphics[scale=0.4]{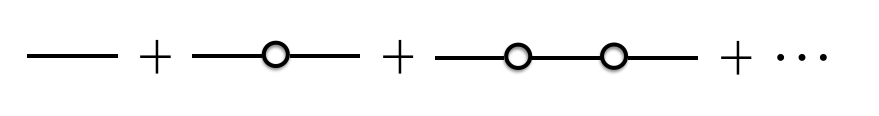}
%\vskip-30pt
\caption{Part of these Feynman diagrams need to resummed to all orders in the coupling.}
\label{fig:chains}
\end{figure}

Although we have only sketched explicitly how to compute the $O(g^2)$ contributions (which however through the $\beta$ functions  \eqref{beta-physical} or \eqref{beta-scaled} furnish higher order contributions and indeed resum these in the usual fashion), we trust the treatment of higher order contributions along these lines is also clear. 

We note that the scalar field theory will also be subject to some corrections that must be handled non-perturbatively in the IR. In particular, classes of Feynman diagrams made by replacing the propagators $\Delta_\Lambda$ by the chain of corrections shown in fig. \ref{fig:chains}, as well as providing higher order $\vp$ interactions, induce a mass $m^2(\Lambda)$. From \eqref{fourier-sol}, and setting $\vp=0$ in \eqref{Vpp} and iterating \eqref{dn-phi}:\footnote{or consulting known formulae for Hermite polynomials}
\be 
\label{m2}
m^2(\Lambda) = -\int^\infty_{-\infty}\!\frac{d\vpi}{2\pi}\, \vpi^2\, \fV_\p(\vpi,\Lambda)\, {\rm e}^{%\exp\left(
-\frac{\vpi^2}{2}\Omega_\Lambda}\ =\ \frac{a}{\sqrt{\pi}\Lambda} \sum^\infty_{n=0} (2n+1)!!\left(-\frac{2a^2}{\Lambda^2}\right)^{\!n+1}\!\!\!\!\!\!g_{2n}(\Lambda)\,.
\ee 
The corresponding $O(\vp^0)$ corrections in fig. \ref{fig:chains} thus appear as a power series in $m^2/p^2$.
If we try to treat these order by order perturbatively in the couplings, when inserted into loop corrections (such as those of figs. \ref{fig:tadpoles} or \ref{fig:melons}) we obtain diagrams of ever increasing divergence as the IR cutoff $\Lambda\to0$. This problem is clearly related to the one we noted at the end of the previous subsection. Instead therefore we need to replace $\Delta_\Lambda(p)$ by $C_\Lambda(p)/(p^2+m^2)$, singling out $m(\Lambda)$ for non-perturbative treatment in the IR. At the same time we should use \eqref{m2} to eliminate one degree of freedom, for example $g_0(\Lambda)$, in favour of $m^2(\Lambda)$ in the equations.

We recognise that the $-\half\, \dd\Lambda{n}\left({\partial}_\mu\vp\right)^2$ operators through the chain of diagrams \ref{fig:chains} similarly induce a wavefunction renormalization. These do not result in the same way in IR divergences. Similarly all higher derivative operators \eqref{derivative-operator-basis-physical} are IR safe in this sense. 

Note that in a correctly formed continuum limit, all contributions from all operators are UV safe and do not need non-perturbative resummation in this regime, apart from using the $\beta$ function to resum the evolution of any marginally relevant coupling. This follows because such a continuum limit depends only on the (marginally) relevant couplings whose scaled versions must vanish in the limit $\Lambda\to\infty$ so that the flow emanates from the Gaussian fixed point as required.

\section{Unitarity and universality}
\label{sec:unitarity}

We are not of course claiming that scalar field theory with wrong sign kinetic term, when considered as a continuum quantum field theory in its own right, is free from physical problems. In Minkowski signature, the wrong sign for the kinetic term implies either a Hamiltonian unbounded from below, or a Fock space with negative norm states (see \eg sec. 8 of \cite{Arnone:2001iy}). Presumably related, the dimensions $[\delta_n]<1$, \cf \eqref{delta-dim}, all violate the unitarity bound. The existence of higher derivative relevant eigenoperators, \cf \eqref{derivative-operator-basis-renormalised}, leads to further concerns for unitarity.
Finally the fact that it is specified by an infinite number of relevant couplings is  phenomenologically \TRM{useless}, and raises questions about universality as already touched on in sec. \ref{sec:non-deriv}. However 
it is natural to expect that these problems disappear when the structure is appropriately embedded into gravity, as discussed in the Introduction and sec. \ref{sec:QG}.

\section{RG evolution on a manifold}
\label{sec:compact}

As we have seen, even at the linearised level, RG evolution %of the eigenoperators 
plays a crucial r\^ole.
%Recall that the continuum limit is parametrised by the couplings conjugate to the (marginally) relevant eigenoperators (\cf sec. \ref{sec:plus}). The eigenoperators themselves therefore play a crucial r\^ole. 
By the quantisation condition, the eigenoperators are given at the bare level by the operators in eqn. \eqref{derivative-operator-basis-physical} with $\Lambda=\Lz$, %with field amplitude dependence that decays exponentially on this scale, 
as given by the coefficient functions \eqref{physical-dnL}. At the linear level these composite operators do not interact with each other, but they nevertheless evolve under lowering the cutoff, by tadpole quantum corrections as in fig. \ref{fig:tadpoles}.
In $\mathbb{R}^4$, by the eigenoperator property, they are form invariant under this evolution, with the inherent scale now equal to the infrared cutoff, as in eqn. \eqref{physical-dnk}, becoming the distributions \eqref{derivative-operator-basis-renormalised} in the physical limit in which the infrared cutoff is removed, \ie as $k\to0$. 

\subsection{Eigenoperators on a manifold}
\label{sec:compact-eigen}

On a (Euclidean) spacetime manifold $\mathcal{M}$ that is not $\mathbb{R}^4$, the bare operators are still the same, because these operators are defined at $\Lz$, the UV scale that is eventually diverging, corresponding to vanishing distances where the spacetime is indistinguishable from $\mathbb{R}^4$. However the quantum corrections are modified at long distances by the spacetime geometry. To be specific it is sufficient to consider the evolution of the potential operator $\dd\Lz{n}$, as defined in \eqref{physical-dnL}, since a general eigenoperator is also made with this term, and  the top part, \eqref{derivative-operator-basis-physical} with covariant derivatives as appropriate, evolves in the same way. 

The evolution will be given by \eqref{tadpoles-evolution}, where the propagation now takes place on the manifold (and thus also a $\sqrt{g}$ is included in the integral over $x$). 
Actually, until we know the form of the full theory of quantum gravity, we do not know for sure what replaces \eqref{tadpoles-evolution} in the general case.\footnote{For example whether $\vp$ is conformally coupled to the background curvature, \cf  sec. \ref{sec:QG}.}
%It seems reasonable to assume that it can be described in terms of the scalar field theory propagating on a manifold described by a background metric, thus subject to background diffeomorphism invariance. It is also natural to assume that it is conformally coupled to the background curvature, see sec. \ref{sec:QG}. 
For the general arguments below we do not need the precise definition, only that it reduces to the flat space version when the background metric $g_{\mu\nu}\to\delta_{\mu\nu}$. Then in the fully worked example we choose the metric to be $\delta_{\mu\nu}$.

On the other hand, since the bare operator is the same, the identity \eqref{bare-delta-Omega} still holds and thus the bare operator can still be expressed as \eqref{bare-delta-prop}, \emph{where the integration is still over} $\mathbb{R}^4$.  Thus combining \eqref{bare-delta-Omega} and \eqref{tadpoles-evolution}, the quantum corrections above $k$ no longer precisely cancel to give \eqref{bare-delta-Omega} with $\Lz$ replaced by $k$, but leave a modified version where:
\be 
\label{delta-M-Omega}
\dd{\!k,\Lz}n = \exp\left(\frac12\,\Omega_{k,\Lz}(x)\, \frac{\partial^2}{\partial\vp^2}\right) \dd{}n\,,
\ee
and
\be 
\label{Omega-M-kL}
\Omega_{k,\Lz}(x) = %\lim_{\Lz\to\infty} \left( 
|\langle \ph(x) \ph(x) \rangle |_{\mathbb{R}^4}-|\langle \ph(x) \ph(x) \rangle |_{\mathcal{M}}\,. %\right)\,.
\ee 
Here the first term is $\Omega_\Lz$, as defined in \eqref{Omega}, while the second term is from propagation on the manifold $\M$ and is regulated by $C^\Lz_k$. 
%It is to be understood that the limit as $\Lz\to\infty$ should be taken. 
In general the second term depends on the position of the point $x$ in $\mathcal{M}$, and thus $\dd{\!k,\Lz}n$ has $x$ dependence through $\Omega_{k,\Lz}(x)$ as well as through its dependence on the field $\vp(x)$.

Consequentially, the operators are no longer form invariant, but pick up ``finite size'' corrections, and will retain some dependence on the UV regularisation while $\Lz$ is finite. However we can expect that $\Omega_{k,\Lz}(x)$ becomes independent of the latter in the limit $\Lz\to\infty$, in particular the operators will again be automatically renormalized, because the tadpole corrections will continue to wipe out all dependence on higher scales providing $k\gg 1/L$, where $L$ is a characteristic length scale for the manifold. 
This will continue to work as $k$ is lowered, until $k$ is comparable to $1/L$,
%Na\"\i vely one would then expect the finite size $L$ to take over the r\^ole of the infrared cutoff so that the evolution effectively freezes out. However as we will see below, something more dramatic happens.
%until $1/k$ becomes comparable with the length dimensions of the manifold, %After this point 
after which the infrared properties should primarily be set by the geometry. 
In particular in the limit that $k\to0$, we expect that $\Omega_{k,\Lz}(x)$ will therefore become a finite universal function of this geometry. We call this function
\be 
\label{Omega-p}
\Omega_\p(x) := \lim_{\Lz\to\infty\atop k\to0} \Omega_{k,\Lz}(x)\,.
\ee
%Thus we expect to find that the IR evolution is effectively frozen out: the operators now only exponentially suppress large field amplitude, with the suppression scale set by the size of the manifold.
%From the above discussion we see that combining \eqref{bare-delta-prop} and \eqref{tadpoles-evolution}, results in 
%%in the physical limit $k\to0$ and $\Lz\to\infty$, we are left with
%\be 
%\label{delta-M-Omega}
%\dd{\!k,\Lz}n = \exp\left(\frac12\,\Omega_{k,\Lz}(x)\, \frac{\partial^2}{\partial\vp^2}\right) \dd{}n\,,
%\ee
%where
%\be 
%\label{Omega-M-kL}
%\Omega_{k,\Lz}(x) = %\lim_{\Lz\to\infty} \left( 
%|\langle \ph(x) \ph(x) \rangle |_{\mathbb{R}^4}-|\langle \ph(x) \ph(x) \rangle |_{\mathcal{M}}\,. %\right)\,.
%\ee 
%Here the first term is $\Omega_\Lz$, as defined in \eqref{Omega}, while the second term is on the manifold $\M$ and is regulated by $C^\Lz_k$. 
%%It is to be understood that the limit as $\Lz\to\infty$ should be taken. 
%In general the second term depends on the position of the point $x$ in $\mathcal{M}$, and thus $\dd{\!k,\Lz}n$ has $x$ dependence through $\Omega_{k,\Lz}(x)$ as well as through its dependence on the field $\vp(x)$. 
By comparing \eqref{bare-delta-Omega} and \eqref{physical-dnL}, we see immediately that evaluating \eqref{delta-M-Omega} gives again the same form for eigenoperators on $\M$ as in \eqref{physical-dnk}, but with $\Omega_k$ replaced by $\Omega_{k,\Lz}(x)$. Taking the limits \eqref{Omega-p} we get the physical eigenoperators $\dd{\p}{n}$, which are thus given by
%In the limits $\Lz\to\infty$ and $k\to0$, we get the physical eigenoperators $\dd{\p}{n}$. These thus also take the form 
%\eqref{physical-dnk} with $\Omega_k$ replaced by $\Omega_\p(x)$.
\be
\label{physical-p}
\dd\p{n} = \frac{\partial^n}{\partial\vp^n}\, \dd\p0\,, \qquad{\rm where}\qquad \dd\p0 = \frac{1}{\sqrt{2\pi\Omega_\p}}\,\exp\left(-\frac{\vp^2}{2\Omega_\p}\right)\,.
\ee
%Again these take the same form as in \eqref{physical-dnk} but with $\Omega_k$ replaced by 
%\be 
%\label{Omega-p}
%\Omega_\p(x) := \lim_{\Lz\to\infty\atop k\to\infty} \Omega_{k,\Lz}(x)\,.
%\ee
Evidently, $\Omega_\p=0$ if the manifold is $\mathbb{R}^4$, and we return to  $\dd{\p}{n}=\dd{}n$.
%From the above discussion, if $\M\ne\mathbb{R}^4$ then we expect to find that $\Omega_\p(x)$ is a finite universal function of its geometry. 
Otherwise, by dimensions 
\be 
\label{shape}
\Omega_\p(x) = \frac{\Sh(x)}{4\pi L^2}\,,
\ee
where $\Sh$ is a (universal) dimensionless `shape' function that can thus only depend on dimensionless characterisations of the manifold (the factor $4\pi$ is included for convenience). Providing $\Sh(x)>0$, $\Omega_\p$ acts to suppress large amplitudes $\vp>1/L$. However as we will see, it is also possible for $\Sh$ to be negative. 

\subsection{General linear RG flows on a manifold}
\label{sec:compact-linear}

In this latter case, the operators $\dd{\!k,\Lz}n$ themselves cease to exist below some positive IR cutoff $k$, being the value where, for some $x$,  $\Omega_{k,\Lz}(x)$  first vanishes and then turns negative.
(Here $\Lz$ can be finite or the  continuum limit, $\Lz\to\infty$, could have been taken.) At this point we get a distribution, namely $\dd{}n$, and attempting to flow below this $k$ will result in the operator turning imaginary, as in \eqref{imaginary-V}. Once more, a full understanding at the linearised level is only gained by switching on infinitely many couplings. Consider again the general solution \eqref{fourier-sol} for the potential. This solution now takes the form
\be  
\label{compact-fourier-sol}
V(\vp,k,\Lz) = \int^\infty_{-\infty}\!\frac{d\vpi}{2\pi}\, \fV_\p(\vpi)\, {\rm e}^{%\exp\left(
-\frac{\vpi^2}{2}\Omega_{k,\Lz}+i\vpi\,\vp} \,, %\right)\,,
\ee
where the choice of bare (relevant) couplings fixes the theory, and in particular determines the amplitude suppression scale $\Lambda_\p$. As before, the above expression is meaningful even when $V\notin\Lmm$. Additionally it remains meaningful even
when the eigenoperators themselves fail to exist, since by \eqref{fourierVplargephi} the integral still converges for large $\vpi$ providing $\Omega_{k,\Lz}(x)>-\Lambda_\p^2/2$ for all $x\in\M$. Taking the limits $\Lz\to\infty$ and $k\to0$, the physical potential is now:
\be  
\label{compact-fourier-phys}
V_\p\left(\vp(x),x\right) = \int^\infty_{-\infty}\!\frac{d\vpi}{2\pi}\, \fV_\p(\vpi)\, {\rm e}^{%\exp\left(
-\frac{\vpi^2}{2}\Omega_\p(x)+i\vpi\,\vp(x)} \,,
\ee
and thus asymptotically for large field:
\be 
\label{compact-large-field}
V_\p\left(\vp(x),x\right) \sim \exp\left(-\frac{\vp^2(x)}{\Lambda_\p^2+2\Omega_\p(x)}\right)\,.
\ee
Thus $\Omega_\p(x)$ modifies the amplitude suppression scale, increasing or decreasing it, depending on the sign.
In particular from \eqref{shape}, the given theory only makes sense on manifolds where\footnote{It might be possible to make sense of the limiting case where $\Omega_\p(x)=-\Lambda_\p^2/2$ %hits that bound 
for some points or subspace in $\M$.} 
\be 
\label{lower-bound}
%\Omega_\p(x) > -\Lambda_\p^2/2
\Sh(x) > -2\pi L^2\Lambda^2_\p\qquad \forall x\in \M\,.
\ee
%$\Omega_\p(x)$ is bounded below\footnote{It might be possible to make sense of the limiting case where $\Omega_\p(x)=-\Lambda_\p^2/2$ %hits that bound 
%for some points or subspace in $\M$.} by $-\Lambda_\p^2/2$. 

Judging from the example below, and confirmed in further examples in ref. \cite{Matt1}, manifolds %$\M=\M_-$ 
where $\Sh(x)$ is somewhere negative, have the characteristic that they have at least one other finite length scale %$L_-$ 
which is sufficiently different, already at the $O(1)$ level, from some appropriately defined average length scale $L$.  For the given theory (\viz choice of couplings) such manifolds must thus be larger than a minimum size  
\be 
\label{Lmin}
L> L_{\rm min}= \frac1{\Lambda_\p}\sqrt{\frac{-\Sh_{\rm min}}{2\pi}}\,,
\ee
where $\Sh_{\rm min}$ is the infimum value over all $x\in\M$.
On the other hand, the larger the characteristic length scale $L$, the more inhomogeneous the manifold (the more negative $\Sh$) is allowed to be. 

Indeed we can rephrase this effect in terms of inhomogeneity. Let $\Sh_{\rm max}>0$ be the maximum (strictly supremum) value for $\Sh_{\rm min}$ over a suitable set of such manifolds $\M$ with the same topology. This is naturally a number of $O(1)$, characteristic of what the theory regards as the most symmetric manifold in the set. Then for a given manifold $\M$, the quantity $\I_\M= \Sh_{\rm max} -  \Sh_{\rm min} >0$ is a universal measure of its inhomogeneity (in the sense of being independent of the details of regularisation). Rephrasing \eqref{Lmin}, the inhomogeneity is bounded above depending on the size of the universe:
\be 
\label{inhomogeneity}
\I_\M < \Sh_{\rm max} + 2\pi L^2\Lambda_\p^2\,.
\ee

Evidently, such behaviour could be very attractive within a complete theory of quantum gravity  (\cf sec. \ref{sec:QG}), although a full, and dynamical, understanding, will have to wait until the non-linear theory is developed. In particular it cries out for application to cosmology. It explains why the initial conditions for inflation had to be sufficiently smooth. It
%lead to an explanation for sufficiently smooth initial conditions, 
possibly requires from quantum gravity alone that the early universe approximates a highly symmetric state such as a de Sitter inflationary phase. The restriction on inhomogeneity is maybe sufficient to forbid eternal inflation. Since (classical) fluctuations are restricted anyway, it maybe does away with the need for inflation altogether.  See \eg refs. \cite{Hollands:2002yb,Kofman:2002cj,Hollands:2002xi,Carroll:2010aj} for  discussions relevant to these ideas. Since it ties the minimum size of the universe to the degree of inhomogeneity, and large amplitude inhomogeneities have appeared only recently in the history of the universe,  it could also explain the infamous ``Why now?'' problem, namely that the energy density of matter (including dark matter)  is now similar in magnitude to the apparent energy density of dark energy deduced from the current acceleration of the universe. Finally, assuming spacetime singularities induce infinite inhomogeneity $\I_\M$, it implies ``cosmic censorship'' and somehow a softening of the causal structure of black holes.

%Clearly this would be an attractive feature which would help complete the theory of inflation. Since the theory requires homogeneity of the whole manifold (not just 

%, for all $x\in \M$. (It might also be possible to make sense of the limiting case where $\Omega_\p(x)\ge-\Lambda_\p^2/2$, for all $x\in \M$.)

\subsection{Eigenoperators on a hyper-torus}
\label{sec:torus-eigen}

We now evaluate $\Omega_\p(x)$ in a simple example, verify that it is universal, and demonstrate that requiring $\Sh>0$ restricts the amount of asymmetry in the manifold. We choose the manifold to be a four-dimensional (untwisted) hyper-torus. Such a manifold is of course not a very realistic representation of our universe. The same effects however also appear for other examples \cite{Matt1}, including cases where the time direction is non-compact. We choose the minimum lengths of the non-contractable loops to be $L_\mu$, %By adapting the derivations in ref. \cite{Hasenfratz:1989pk}, we will show that $\Omega_\p$ is a universal function of these four moduli. 
and choose flat coordinates such that $g_{\mu\nu}=\delta_{\mu\nu}$.
In this case 
\be 
\label{torus-tadpole}
|\langle \ph(x) \ph(x) \rangle |_{\mathcal{M}} = \frac1V \sum_{n\ne0} \frac{C^\Lz_k(p_n)}{p^2_n}\,,
\ee
where $p^\mu_n = 2\pi n_\mu/L_\mu$ (no summation over $\mu$), the sum is over all vectors of integers $n\in \mathbb{Z}^4\backslash \{0\}$, %where we exclude the zero mode, 
and  $V=\Pi_{\mu=1}^4 L_\mu$ is the volume of the hyper-torus. 
%and $m$ is a small regulator mass (to be taken to zero when safe to do so). 
Note that since the hypertorus has translation invariance, in this case there is actually no $x$ dependence. Then $\Sh$ can only depend on ratios of length scales.

Also note that since this is a manifold of finite volume, the constant mode (\aka zero mode) $\vp(x)=\vp_0$ is normalizable. It needs to be divided out from the functional measure since a pure kinetic term, and thus the integrand of the partition function at the Gaussian fixed point, does not depend on this (recall related comments at the beginning of sec. \ref{sec:general}). This is the reason for excluding $n=0$ from the sum in \eqref{torus-tadpole}, making it manifestly IR finite. %as is to be expected on general grounds.
 Therefore the limit $k\to0$ in \eqref{Omega-p} can be safely taken, and $\Omega_\p$ is clearly independent of the choice of IR regularisation.

With the infrared cutoff $k>0$ in place, the $n=0$ contribution is not singular. Indeed
\be
\label{limit1} 
\lim_{p\to0} \frac{C^\Lz_k(p)}{p^2} = C'(0)\left(\frac1{\Lzp{2}}-\frac1{k^2}\right)\,,
\ee
where we have used \eqref{sum-rule} and below \eqref{DeltaUV}. Using this to add back the $n=0$ contribution, we can then employ the Poisson summation formula to write \eqref{torus-tadpole} as a sum over winding numbers:
%This mode can be excluded by fiat as indicated above or automatically by using cutoff profiles $C(p^2/\Lambda^2)$ that satisfy $C'(0)=0$, \cf \eqref{sum-rule} and below \eqref{DeltaUV}. Since the calculation is slightly neater in the latter case, we will henceforth in this section specialise to profiles satisfying $C'(0)=0$ and thus take the sum above to be over all $n\in \mathbb{Z}^4$
%Using the Poisson summation formula, we then write \eqref{torus-tadpole} as a sum over winding numbers
\be 
|\langle \ph(x) \ph(x) \rangle |_{\mathcal{M}} = \int \frac{d^4p}{(2\pi)^4}  \frac{C^\Lz_k(p)}{p^2} \sum_n {\rm e}^{il_n\cdot p}\ -\frac{C'(0)}{V}\left(\frac1{\Lzp{2}}-\frac1{k^2}\right)\,,
\ee
where $l_{n\,\mu} = L_\mu n_\mu$ (not summed over $\mu$) and $n\in \mathbb{Z}^4$ are now the winding numbers. Using \eqref{sum-rule} and \eqref{Omega}, we see that the zero winding number sector, \ie $n=0$, yields the $\mathbb{R}^4$-quantity $\Omega_\Lz-\Omega_k$, and thus from \eqref{Omega-M-kL} we find that
\be 
\label{step1}
\Omega_{k,\Lz}\ =\ \ \Omega_k +\frac{C'(0)}{V}\left(\frac1{\Lzp{2}}-\frac1{k^2}\right) 
 -\int \frac{d^4p}{(2\pi)^4}  \frac{C^\Lz_k(p)}{p^2} \sum_{n\ne0} {\rm e}^{il_n\cdot p}\,.
%= - \sum_{n\ne0} |\langle \ph(0) \ph(l) \rangle |_{\mathbb{R}^4}
\ee
Since the last term is a sum of propagators to separated points, we see that $\Omega_{k,\Lz}$ is manifestly UV finite, as we already argued above on general grounds. We can therefore safely take the limit $\Lz\to\infty$, with the result clearly independent of the method UV regularisation (in this case the UV cutoff profile). As we have already seen that it is IR safe, we have thus proved that $\Omega_\p$ is well-defined and universal, as we claimed. 

We are free to choose the IR cutoff profile to facilitate the remaining calculation. We set $C(p^2/k^2)= {\rm e}^{-p^2/k^2}$.\footnote{For a different choice see ref. \cite{Hasenfratz:1989pk}; we otherwise essentially follow their derivation.} Recall that by \eqref{sum-rule}, $C_k(p)=1-C(p^2/k^2)$. Taking limits where it is safe to do so, we can thus write:
\be 
\Omega_\p = \frac1{Vk^2}-\int \frac{d^4p}{(2\pi)^4}\, \int^{1/k^2}_0\!\!\!\!\!\!\!\!\!\!d\alpha\,\,{\rm e}^{-\alpha p^2} \sum_{n\ne0} {\rm e}^{il_n\cdot p}\,,
\ee
where we have expressed the IR cutoff through a Schwinger parameter, and the $k\to0$ limit should hereafter be understood.
Performing the momentum integral, and substituting $\alpha = L^2 t/4\pi$, where $L = V^{1/4}$ is the geometric mean of the $L_\mu$, gives 
\be 
\label{stepTheta}
\Omega_\p  = \frac1{Vk^2} -\frac1{4\pi L^2}\int_0^{\frac{4\pi}{L^2k^2}} \frac{dt}{t^2} \left[ \Pi_{\mu=1}^4\, \Theta\left(\frac{L^2_\mu}{tL^2}\right)-1\right]\,,
\ee
where we have introduced the third Jacobi theta function (at Jacobi $\nu=0$, $x>0$):
\be 
\Theta(x) := \sum_{n=-\infty}^\infty\!\! {\rm e}^{-\pi n^2 x}\,.
\ee
Splitting the integral into two pieces about $t=1$, the first piece is given by $s(L_\mu/L)$ where
\be 
\label{s}
s(\ell_\mu) := \int_0^1 \frac{dt}{t^2} \bigg( \Pi_{\mu=1}^4\, \Theta\left({\ell^2_\mu}/{t}\right)-1\bigg)\,.
\ee
In the $t\ge1$ part we substitute $t\mapsto1/t$ and use the identity $\Theta(x) = (1/\sqrt{x})\, \Theta(1/x)$ (which straightfowardly follows from a further application of Poisson resummation) to cast it in terms of the above function plus a remainder. The latter
%\be 
%-\frac1{4\pi L^2}\int^1_{\frac{L^2k^2}{4\pi}}  \frac{dt}{t^2} \left(1-t^2\right)\,,
%\ee
in particular cancels the explicit IR divergence in \eqref{stepTheta}. Thus finally, using \eqref{shape}, we find 
\be 
\Omega_\p = \frac{\Sh(L_\mu/L)}{4\pi\sqrt{V}}\qquad{\rm where}\qquad \Sh(\ell_\mu) := 2-s(\ell_\mu)-s(1/\ell_\mu)\,.
\ee
By dimensions, $\Sh$ only depends on the ratios $L_\mu/L$. Symmetry under permutation of the $L_\mu$ follows from the symmetries of the torus. However we note further that $\Omega_\p$ and $\Sh$ are invariant under the simultaneous inversion of all moduli: $L_\mu \mapsto L^2/L_\mu$ (which also preserves the overall volume $V$). It can be extended to a larger group involving the modular group and twisted torii. This intriguing symmetry is reminiscent of T-duality in String Theory \cite{Green:1982sw,Kikkawa:1984cp,Sakai:1985cs},
except that there radii are inverted using the string scale $\alpha'$, whereas here the scale is set by the manifold itself. Again a comprehensive understanding of its significance in the current context will have to await the development of the full quantum gravity. 

At the symmetric point where all $L_\mu = L$, we find numerically that $\Sh\equiv \Sh_{\rm max}=1.765$, in agreement with ref. \cite{Hasenfratz:1989pk}, and confirming the general expectation that $\Sh_{\rm max}$ is a number of $O(1)$. On the other hand $\Sh$ vanishes already if for example:
\begin{enumerate}
\item[(a)] $L_1 = 2.709\, L$ with the other three $L_\mu$ equal (thus to $0.7173\, L$),
\item[(b)] thus also the dual version $L_1 = 0.3691\, L$ and the other three $L_\mu = 1.394\, L$, 
\item[(c)] $L_1 = L_2 = 2.457\, L$ with the other pair $L_3=L_4=0.4069\, L$. 
\item[(d)] $L_\mu = 1.487\, L_{\mu+1}$\ ($\mu=1,2,3$).
\end{enumerate}
(Combined with permutation symmetry, (c) and (d) are self-dual.) With the $L_\mu$  further apart, these configurations result in $\Sh<0$, which implies a minimum allowed size for such a manifold, for example from
\eqref{lower-bound} we can write this in terms of the
 space-time volume as:
% \cf below \eqref{compact-large-field}:
\be 
V > \frac{\Sh^2(L_\mu/L)}{4\pi^2\Lambda^4_\p}\,.
\ee

%It is sufficient to consider the evolution of the potential operator $\dd\Lz{n}$, as defined in \eqref{physical-dnL}, since a general eigenoperator is made with this term, and  the top part, \eqref{derivative-operator-basis-physical},  evolves in the same way. On a general manifold, the evolution is given by \eqref{tadpoles-evolution}, except that the propagation and the integration now takes place on the manifold, with involvement of the metric $g_{\mu\nu}$ in the usual way. On the other hand, since the bare operator is the same, the identity \eqref{bare-delta-Omega} still holds and thus the bare operator can still be expressed as \eqref{bare-delta-prop} where \emph{the integration is over $\mathbb{R}^4$ however}.

\section{Implications for quantum gravity}
\label{sec:QG}

The discoveries we have reported in this paper point towards gravity being after all a perturbatively renormalizable quantum field theory, albeit of a new and dramatically different kind. Of course physical processes are described by working with the theory in Minkowski signature, or by using some continuation appropriately adapted to the process at hand (see \eg the recent discussion \cite{Feldbrugge:2017mbc}). However before such processes can be investigated, one must actually construct such a theory. To do this we need to formulate it
%in terms of 
%%See pp180, 217, 220, 223, 225-8 & 243-5.
%As we have already noted, in order to understand gravity 
in Wilsonian terms, which means that we need to study its fluctuations around Euclidean $\mathbb{R}^4$ (see secs. \ref{sec:Intro} and \ref{sec:plus}). Then, reflecting the unboundedness of the Euclidean signature action, the conformal factor  has the wrong sign kinetic term.  Considered on its own, we have shown in the previous sections how to make sense of its Wilsonian RG behaviour,
%at the linearised level, and in perturbation theory, 
uncovering novel and promising properties (further explored in ref.  \cite{Matt1}).
%We have seen that the eigenoperators are non-perturbative in $\hbar$. The continuum limit is built using the relevant eigenoperators, each one is associated with a relevant coupling. 
Now we discuss %what we can deduce about 
%how %the conformal factor field must be put back with 
%this must be merged with the rest of gravity to obtain a consistent quantization. 
what this implies for the full theory of quantum gravity.

The key observation from the Wilsonian RG, is that the continuum theory can be constructed if %for an appropriate parametrisation, 
the scaled bare action in the limit $\Lz\to\infty$ is just the Gaussian fixed point plus a vanishing perturbation which is the linearised interaction expanded only over (marginally) relevant eigen-operators. This provides the boundary condition for the renormalized trajectory, and renormalizability can then be expected to follow provided that all bare relevant couplings are included that are induced by requiring finite couplings at physical scales. 
%In other words we expect that all bare relevant couplings are non-vanishing unless some symmetry principle allows to set some to zero.
%with the conjugate couplings evolving as required.
%renormalizability \emph{will be guaranteed} if we can build the theory around the Gaussian fixed point using \emph{all and only} the relevant eigen-operators. 
%Indeed renormalizability is then guaranteed since it is nothing but a restatement of the resulting universality of the renormalized trajectory emanating from the fixed point. 
More generally, if bare irrelevant couplings are needed, they must stay close enough to the Gaussian fixed point to remain within its domain of attraction.  Just as discussed in sec. \ref{sec:plus}, we can then anticipate that their dimensionful values must  
actually vanish in the limit as $\Lz\to\infty$.
%This provides us with an infinite set of constraints that must be satisfied for a valid continuum limit. 
%We can anticipate that infinitely many bare relevant couplings will be required in order to ensure a finite 
%for the continuum limit. Switching off any linear combination, would require some special fine tuning, which can only be justified by the imposition of some symmetry which is preserved by the regularisation. 

For the conformal factor on its own, this means in particular that the bare theory must sit inside $\Lmm$, using %all and only 
the relevant interactions of the form \eqref{derivative-operator-basis-physical}. Since these eigenoperators are non-perturbative in $\hbar$, quantum gravity must also be non-perturbative in $\hbar$. %It does not make sense to work at a fixed number of loops.
%The loop expansion is meaningless here. 
Therefore we cannot organise contributions by the loop expansion,
however calculations can proceed %by working 
perturbatively in $\kappa$ (\ie Newton's coupling \cf sec. \ref{sec:Intro}). Since the traceless fluctuation $\h_{\mu\nu}$ has the right sign for its kinetic term, \cf \eqref{Gaussian},   eigen-operators involving only $\h_{\mu\nu}$  are built in $\Lm+$, \ie are polynomials of $\h_{\mu\nu}$ and its space-time derivatives, generalising sec. \ref{sec:plus} (see also sec. \ref{sec:derivative-ops}). In particular  $[\h_{\mu\nu}]=1$ and $\tilde{h}_{\mu\nu} = h_{\mu\nu}/\Lambda$,
as follows from the canonically normalized kinetic term \eqref{Gaussian}, and the Hilbert space $\Lm+$ is defined through the norm ${\rm e}^{-a^2\tilde{h}_{\mu\nu}^2}$.
Extending sec. \ref{sec:derivative-ops},  it is  thus clear that the general eigenoperator 
% (mixed) involving both $\vp$ and $\h_{\mu\nu}$, 
is built using a top term 
%which is either
\be 
\label{top}
\dd\Lambda{n} \,\sigma(\h,\partial,\partial\vp)\,,
\ee
%or $\sigma(\h,\partial)$, 
where $\sigma(\h,\partial,\partial\vp)$ is a Lorentz invariant monomial involving some or all of the components indicated (and thus $\h_{\mu\nu}$ can appear here differentiated or undifferentiated or not at all). These perturbations form the Hilbert space ``$\Lm{}$'' %at the intersection $\Lm{} = \Lm+\cap\Lmm$, which is to say that they 
of interactions that are square integrable under ${\rm e}^{a^2(\tp^2-\tilde{h}_{\mu\nu}^2)}$. Clearly this includes the $\vp$ eigen-perturbations that are purely in $\Lmm$, since these interactions are still square-integrable under the new measure. But $h_{\mu\nu}$
eigen-perturbations that are purely in $\Lm+$ are not allowed since they are not square integrable under the new measure (there is nothing to mitigate the ${\rm e}^{a^2\tp^2}$ part). If we included such interactions we would destroy the $\vp$ part of the Hilbert space structure and as we will see, also renormalizability.
%In practice we cannot use this latter pure-$\h$ type of eigenoperator since it will induce non-vanishing irrelevant couplings in the bare action, the original problematic terms \eqref{irrelevant} being examples. 
The scaling dimensions of the eigenoperators are the ones expected at the Gaussian fixed point, in particular if %the mass dimension 
$[\sigma(\h,\partial,\partial\vp)]=d_\sigma$, then 
the scaling dimension of the full eigenoperator is 
%the one expected at the Gaussian fixed point, namely 
$d_\sigma+[\delta_n] = d_\sigma-1-n$.

It is tempting to assume that all symmetries are preserved and that we can discuss the issue within the framework of a classical action. But neither of these assumptions is true: the regularisation (and not only this as we will discuss) breaks or at least deforms local symmetries, and thanks to the conformal factor, the action is never classical but always non-perturbatively quantum. The usual arguments proceed by assuming diffeomorphism invariance, leading at the classical level to a series of interactions \eqref{irrelevant} organised by powers of $\kappa$, after which quantum corrections can be analysed. Here the interactions at each new power of $\kappa$ arise simultaneously from both directions: on the one hand from the quantum corrections induced by interactions with a lower power of $\kappa$, and on the other hand by the constraints of the quantum (BRST) version of diffeomorphism invariance. %imposed on the physical action. 

Provided the latter at least incorporates %for the physical effective action (through BRS symmetry) 
the linearised diffeomorphism invariance enjoyed by \eqref{EHbilinear}, and that the kinetic term remains second order in derivatives at the bare level, back in Minkowski signature this is a theory of gravitons with just two transverse polarisations. In particular this also ensures that in Minkowski signature, the conformal mode is non-dynamical, and thus that the wrong-sign kinetic term does not lead to a break-down of unitarity. 

To the extent that the low energy effective description can be assumed to be classical, many related arguments of consistency then effectively enforce that it coincides with General Relativity \cite{Gupta:1954zz,Kraichnan:1955zz,Feynman:1996kb,Weinberg:1965rz,Ogievetsky:1965,Wyss:1965,Deser:1969wk,Boulware:1974sr,Fang:1978rc,Wald:1986bj,Boulanger:2000rq}. Given all the experimental tests, this seems surely to be required phenomenologically.  As we have been emphasising however, according to the theory we are uncovering, gravity must in reality be non-perturbatively quantum at all scales. This aspect lies at the heart of the restrictions on inhomogeneity, which as discussed in sec. \ref{sec:compact-linear}, themselves look so promising phenomenologically. We can add that the  tendency to IR divergence at the interacting level (see the end of sec. \ref{sec:pert2-flow}) make it tempting to speculate that gravitational dynamics will receive important corrections at large scales, raising the prospect that these effects could be ones attributed to dark matter, and perhaps even have a r\^ole in explaining conflicting experimental measurements of Newton's coupling \cite{Mohr:2015ccw}. Clearly there is some tension with the conclusion we reached at the beginning of this paragraph. The actual extent to which General Relativity is modified will only be revealed once the full theory is developed.

Since the BRST invariance is broken by our regularisation, bare operators corresponding to its breaking, will have non-vanishing couplings, even though the corresponding physical expressions are tuned to vanish. To avoid the breaking of this quantum version of diffeomorphism invariance, one might hope to reformulate the arguments using dimensional regularisation. However, since quadratic divergences of a massless field are crucial to the definition of the $\vp$ eigenoperators, dimensional regularisation would appear to be inapplicable. In principle we could try to finesse the difficulties by basing the formulation on the fact that $\Omega_\p$ in \eqref{Omega-p} is actually independent of regularisation and thus also the physical operators \eqref{physical-p} are independent of regularisation. But to discuss renormalizability we need access to the bare operator, which requires using only the first term in \eqref{Omega-M-kL}. This vanishes in dimensional regularisation, which by \eqref{physical-dnL} implies that all the bare operators also vanish. 
We could try the usual expedient of adding a mass term for $\vp$ by hand. However adding a mass term breaks the realisation of diffeomorphism invariance we were trying to preserve, meaning that we appear to be no better off than with the rigorously more secure regularisation scheme we are currently using.

We need to avoid being forced by the parametrisation, equivalently the realisation of diffeomorphism invariance, to include irrelevant operators with corresponding non-vanishing couplings in the limit $\Lz\to\infty$ (this being the usual problem). To gain some feeling for the parametrisation required, let us imagine for the moment that the theory can be constructed by starting from a diffeomorphism invariant classical action. Then since the action will be \eqref{EH}, and the kinetic terms have to appear explicitly as in \eqref{Gaussian}, any parametrisation can be reduced to the question of how
to parametrise the metric $g_{\mu\nu}$. To linear order in the fields we know already that this takes the form \eqref{param-perturbative}, in order to obtain \eqref{Gaussian} after using the Feynman -- De Donder gauge \eqref{Feynman-DeDonder}. This suggests writing
\be 
\label{metric-overparam}
g_{\mu\nu} = \left(1+\frac\kappa4\,\vp\right)^2  \hat{g}_{\mu\nu}\,,
\ee
so that \eqref{EH} becomes:
\be 
\label{EH-phi}
\cL_{EH} =   -\frac34 \sqrt{\hat{g}}\,\hat{g}^{\mu\nu}\partial_\mu\vp\partial_\nu\vp 
-\frac2{\kappa^2}\,\sqrt{\hat{g}}\hat{R}\left(1+\frac\kappa4\,\vp\right)^2\,.
\ee
If $\hat{g}_{\mu\nu}=\delta_{\mu\nu}$, this gives us the required kinetic term for $\vp$ (before getting $\frac14(\partial\vp)^2$ from gauge fixing) and nothing else.
From \eqref{param-perturbative} we then know that to linear order in the fields,  $\hat{g}_{\mu\nu} = \delta_{\mu\nu} + \kappa\,\h_{\mu\nu}$.
But such an unadorned $\hmn$ will lead us straight back into the space of non-renormalizable finite irrelevant interactions \eqref{irrelevant}, and take us outside $\Lm{}$. Instead we need to 
%use the terms in \eqref{top}, which thus stay in $\Lm{}$ and can provide sufficient protection for higher order $\hmn$ interactions. 
protect it by using the $\vp$ operators \eqref{delta}.
For example we could try replacing $h_{\mu\nu}$ with the marginal operator $\dd\Lambda0\,h_{\mu\nu}$,  or with $\dd\Lambda{n}\,h_{\mu\nu}$ for some $n>0$, which is a relevant operator. On the other hand once we use one such a basis operator, perturbative  quantum corrections (\ie in $\kappa$, non-perturbative in $\hbar$) will generate infinitely many others via \eqref{prod}. 
%By the equivalence theorem \cite{Bergere:1975tr,Itzykson:1980rh} such terms could be reparametrised away
Thus to renormalize the theory we expect to need to extend this to an infinite sum over such operators, so we are led to try $\hat{g}_{\mu\nu} = \delta_{\mu\nu} + \kappa\,f_{1}\h_{\mu\nu}$, where $f_{1}(\vp,\Lz)\in\Lmm$ is a general \emph{coefficient function}. Thus the general structure described in sec. \ref{sec:general} can be expected: the effective interaction will be in $\Lm{}$ at cutoff scales $\Lambda$ higher than some $\Lz$, leaving $\Lm{}$ at some $\Lp$; with further care, complete flows exist, leading to the inhomogeneity effects discussed in sec. \ref{sec:compact}.
%with careful further choice of the couplings, the flow exists all the way down to the physical limit $\Lambda\to0$.

Substituting such an expansion into \eqref{EH-phi} will lead to higher order $\hmn$ interactions, with $\vp$-dependent coefficients that can be expanded over the $\dd\Lambda{n}$ basis using \eqref{prod}. At this point we have to face the fact, as we saw in eqn. \eqref{prod-evolved}, that the flow even at the linearised level does not respect the product structure, and thus here does not respect 
the fact that these operators came from some power of (differentials of) $f_1$. This will be true even if we were able to construct a diffeomorphism invariant flow equation \cite{Morris:2016nda}.
The only way we can match the result to $\hat{g}_{\mu\nu}$ at some other scale,
%Since each operator in the effective action evolves in a way already at the linearised level that does not respect that it came from some power of the coefficient funciton, that would only be possible if we allow $\hat{g}_{\mu\nu}$ 
is to give the latter sufficiently many parameters to reproduce the result of this evolution. We are thus led
to consider very general expansions, schematically (derivative operators might also be needed)  
\be 
\label{g-exp}
\hat{g}_{\mu\nu} = \delta_{\mu\nu} 
+\kappa f_{1}\, \h_{\mu\nu} +\kappa^2 f_{2}\, \h^{\ \alpha}_\mu \h_{\alpha\nu} +\cdots\,,
\ee
each operator with their own coefficient function $f_i(\vp,\Lambda)$. Substituting this expansion into \eqref{EH-phi}, it is clear that this can come from a bare level action where all the interactions are of form \eqref{top}, in particular cubic and higher $\hmn$ interactions appear together with their `protection' via $\vp$ interactions in $\Lmm$. Indeed since $\hat{R}$ vanishes for flat $\hat{g}_{\mu\nu}$, it is reconstructed from interactions all of which contain at least one coefficient function. Then the observations in sec. \ref{sec:perturbation-th} apply. Thus $\partial_\mu f_{j} =\partial_\vp f_{j}\, \partial_\mu\vp$ is in $\Lmm$ by \eqref{d-delta}, products of the $f_{j}$ are in $\Lmm$ by \eqref{prod}, and the explicit instances of $\vp$ in the last term in \eqref{EH-phi} are absorbed into $\Lmm$ by \eqref{dn-phi}. 
We  thus see that the r\^ole of the infinite number of relevant couplings in the conformal sector, \cf \eqref{delta} and sec. \ref{sec:derivative-ops}, is to allow for such a sufficiently general parametrisation. 

So far we have only discussed what happens when we aim for the Einstein-Hilbert action \eqref{EH}. With infinitely many relevant directions of arbitrarily high dimension, one should worry that covariant higher derivative contributions could also be relevant. In particular ones which have an $O(h^2)$ piece, that can for example come from $g_{s} R^2/\kappa^2$ (where $g_s$ is its coupling) and the other squared curvatures, are dangerous since they can destroy unitarity by introducing poles of the wrong sign into the propagator \cite{Stelle:1976gc}. In fact the dimensions \eqref{delta-dim} are just right to ensure that this does not happen!
From \eqref{g-exp} such terms look like $g_sf_1^2 h\partial^mh$ for $m\ge4$. For the generic $f_1$ which we are anyway forced to have, such a term contains $\dd\Lambda0\, h\partial^mh$ which is an irrelevant operator of dimension $m+1\ge5$. Thus the corresponding couplings $[g_s]\le-1$, must be set to vanish in the continuum limit. In essentially the same way, one shows that none of the covariant higher derivative operators can be associated with their own bare couplings. 
%That despite the existence of infinitely many towers of relevant operators, these couplings $[g_s]\le-1$ work out all to be irrelevant (and that we furthermore have the freedom to set their bare values to zero), provides further evidence that we are on the right track.

From \eqref{metric-overparam} and \eqref{g-exp} we would deduce that a cosmological constant term is not allowed, since it leads to non-vanishing $\vp$ and $\vp^2$ terms. These operators are not in $\Lmm$ so do not appear at the bare level, and cannot be generated from products of operators that start in $\Lm{}$. Such a conclusion would be clearly attractive, especially given that the theory already has the potential to explain the current acceleration of the universe (\cf sec. \ref{sec:compact-linear}).
However at this point we have to confess to a flaw in these arguments. Nevertheless they show how these structures are important for quantum gravity, and the flaw indicates the path we have to take. 

The problem is that substituting \eqref{g-exp} does not (after appropriate modification of the Feynman -- De Donder gauge fixing) give the kinetic terms \eqref{Gaussian} plus interactions in $\Lm{}$, because the $\hmn$ kinetic term also gets multiplied by $f_1^2$. Writing it as \eqref{Gaussian} plus the interaction
\be 
\label{killer}
\frac12 (f_1^2-1) \left(\partial_\lambda \h_{\mu\nu}\right)^2\,,
\ee
makes this look harmless, particularly if we can arrange for $f_1|_{\vp=0}=1$ so that it is genuinely only interactions. However \eqref{killer} is not in $\Lm{}$. Although the unprotected $(\partial h)^2$ is marginal (thus perturbatively renormalizable), the Hilbert space structure is destroyed and with it the guarantee that quantum corrections are also in $\Lm{}$ (at sufficiently high scales). Indeed \eqref{killer} together with the other $O(h^2)$ interactions when strung together as in fig. \ref{fig:chains} and inserted into Feynman diagrams made using the other interactions, cancel the $f_1$ appearances in internal legs. In fact all the $f_i$ cancel inside loops. Despite the novel context, the equivalence theorem still applies \cite{Bergere:1975tr,Itzykson:1980rh}. Reparametrising the metric does not help, cosmological constant terms are after all generated, and gravity is still non-renormalizable -- with the same structure of divergences. 

The root cause of the failure is where we flagged it be, in the paragraphs above \eqref{metric-overparam}. We cannot start from a diffeomorphism invariant classical action. Instead we must go directly to a quantum action subject to some quantum version of diffeomorphism invariance. The known consistency constraints \cite{Gupta:1954zz,Kraichnan:1955zz,Feynman:1996kb,Weinberg:1965rz,Ogievetsky:1965,Wyss:1965,Deser:1969wk,Boulware:1974sr,Fang:1978rc,Wald:1986bj,Boulanger:2000rq} appear at first sight to leave no room for an alternative quantum theory. However all of these works assume one or more properties, in particular justified by the assumed existence of a classical limit,  that either now do not apply or become significantly softened.

\bigskip\bigskip

\section*{Acknowledgments}
It is a pleasure to thank Chris Sachrajda for helpful conversations about finite size effects, and Matt Kellett for helpful discussions stemming from the further examples of $\Omega_\p$ \cite{Matt1}. I acknowledge support from both the Leverhulme Trust and the Royal Society as a Royal Society Leverhulme Trust Senior Research Fellow, and from STFC through Consolidated Grants ST/L000296/1 and ST/P000711/1.

\vfill
\newpage 

%\appendix

%\newpage
%\mbox{}
%\newpage

\bibliographystyle{hunsrt}
\bibliography{references} %%from now on (14/7/15) this is the global references list!

\begin{thebibliography}{10}

\bibitem{tHooft:1974toh}
Gerard 't~Hooft and M.~J.~G. Veltman.
\newblock {One loop divergencies in the theory of gravitation}.
\newblock {\em Ann. Inst. H. Poincare Phys. Theor.}, A20:69--94, 1974.

\bibitem{Goroff:1985sz}
Marc~H. Goroff and Augusto Sagnotti.
\newblock {Quantum Gravity at Two Loops}.
\newblock {\em Phys. Lett.}, B160:81--86, 1985.

\bibitem{Goroff:1985th}
Marc~H. Goroff and Augusto Sagnotti.
\newblock {The Ultraviolet Behavior of Einstein Gravity}.
\newblock {\em Nucl. Phys.}, B266:709--736, 1986.

\bibitem{vandeVen:1991gw}
Anton E.~M. van~de Ven.
\newblock {Two loop quantum gravity}.
\newblock {\em Nucl. Phys.}, B378:309--366, 1992.

\bibitem{Weinberg:1980}
S.~Weinberg.
\newblock {Ultraviolet Divergences In Quantum Theories Of Gravitation}.
\newblock {\em In Hawking, S.W., Israel, W.: General Relativity; Cambridge
  University Press}, pages 790--831, 1980.

\bibitem{Reuter:1996}
M.~Reuter.
\newblock {Nonperturbative evolution equation for quantum gravity}.
\newblock {\em Phys.Rev.}, D57:971--985, 1998, hep-th/9605030.

\bibitem{Wilson:1973}
K.G. Wilson and John~B. Kogut.
\newblock {The Renormalization group and the epsilon expansion}.
\newblock {\em Phys.Rept.}, 12:75--200, 1974.

\bibitem{Morris:1998}
Tim~R. Morris.
\newblock {Elements of the continuous renormalization group}.
\newblock {\em Prog.Theor.Phys.Suppl.}, 131:395--414, 1998, hep-th/9802039.

\bibitem{Gibbons:1978ac}
G.W. Gibbons, S.W. Hawking, and M.J. Perry.
\newblock {Path Integrals and the Indefiniteness of the Gravitational Action}.
\newblock {\em Nucl.Phys.}, B138:141, 1978.

\bibitem{Morris:1996nx}
Tim~R. Morris.
\newblock {On the fixed point structure of scalar fields}.
\newblock {\em Phys. Rev. Lett.}, 77:1658, 1996, hep-th/9601128.

\bibitem{Morris:1996xq}
Tim~R. Morris.
\newblock {Three-dimensional massive scalar field theory and the derivative
  expansion of the renormalization group}.
\newblock {\em Nucl.Phys.}, B495:477--504, 1997, hep-th/9612117.

\bibitem{Bridle:2016nsu}
I.~Hamzaan~Bridle and Tim~R. Morris.
\newblock {Fate of nonpolynomial interactions in scalar field theory}.
\newblock {\em Phys. Rev.}, D94:065040, 2016, 1605.06075.

\bibitem{Dietz:2016gzg}
Juergen~A. Dietz, Tim~R. Morris, and Zoe~H. Slade.
\newblock {Fixed point structure of the conformal factor field in quantum
  gravity}.
\newblock {\em Phys. Rev.}, D94(12):124014, 2016, 1605.07636.

\bibitem{Bonanno:2012dg}
Alfio Bonanno and Filippo Guarnieri.
\newblock {Universality and Symmetry Breaking in Conformally Reduced Quantum
  Gravity}.
\newblock {\em Phys.Rev.}, D86:105027, 2012, 1206.6531.

\bibitem{Bollini:1973wu}
C.~G. Bollini and J.~J. Giambiagi.
\newblock {Evanescent couplings and compensation of Adler anomaly}.
\newblock {\em Acta Phys. Austriaca}, 38:211--215, 1973.

\bibitem{Morris:1993}
Tim~R. Morris.
\newblock {The Exact renormalization group and approximate solutions}.
\newblock {\em Int.J.Mod.Phys.}, A 09:2411--2450, 1994, hep-ph/9308265.

\bibitem{Polchinski:1983gv}
Joseph Polchinski.
\newblock {Renormalization and Effective Lagrangians}.
\newblock {\em Nucl.Phys.}, B231:269--295, 1984.

\bibitem{Morris:2015oca}
Tim~R. Morris and Zo{\"e}~H. Slade.
\newblock {Solutions to the reconstruction problem in asymptotic safety}.
\newblock {\em JHEP}, 11:094, 2015, 1507.08657.

\bibitem{Nicoll1977}
J.~F. Nicoll and T.~S. Chang.
\newblock {An Exact One Particle Irreducible Renormalization Group Generator
  for Critical Phenomena}.
\newblock {\em Phys. Lett.}, A62:287--289, 1977.

\bibitem{Bonini:1992vh}
M.~Bonini, M.~D'Attanasio, and G.~Marchesini.
\newblock {Perturbative renormalization and infrared finiteness in the Wilson
  renormalization group: The Massless scalar case}.
\newblock {\em Nucl. Phys.}, B409:441--464, 1993, hep-th/9301114.

\bibitem{Wetterich:1992}
Christof Wetterich.
\newblock {Exact evolution equation for the effective potential}.
\newblock {\em Phys.Lett.}, B301:90--94, 1993.

\bibitem{Keller:1990ej}
G.~Keller, Christoph Kopper, and M.~Salmhofer.
\newblock {Perturbative renormalization and effective Lagrangians in phi**4 in
  four-dimensions}.
\newblock {\em Helv. Phys. Acta}, 65:32--52, 1992.

\bibitem{HHOrig}
Kenneth Halpern and Kerson Huang.
\newblock Fixed-point structure of scalar fields.
\newblock {\em Phys. Rev. Lett.}, 74:3526--3529, May 1995.

\bibitem{Belavin:1975fg}
A.~A. Belavin, Alexander~M. Polyakov, A.~S. Schwartz, and {\relax Yu}.~S.
  Tyupkin.
\newblock {Pseudoparticle Solutions of the Yang-Mills Equations}.
\newblock {\em Phys. Lett.}, B59:85--87, 1975.

\bibitem{tHooft:1976snw}
Gerard 't~Hooft.
\newblock {Computation of the Quantum Effects Due to a Four-Dimensional
  Pseudoparticle}.
\newblock {\em Phys. Rev.}, D14:3432--3450, 1976.
\newblock [Erratum: Phys. Rev.D18,2199(1978)].

\bibitem{tHooft:1977xjm}
Gerard 't~Hooft.
\newblock {Can We Make Sense Out of Quantum Chromodynamics?}
\newblock {\em Subnucl. Ser.}, 15:943, 1979.

\bibitem{Jackiw:1974cv}
R.~Jackiw.
\newblock {Functional evaluation of the effective potential}.
\newblock {\em Phys. Rev.}, D9:1686, 1974.

\bibitem{Nielsen:1975fs}
N.~K. Nielsen.
\newblock {On the Gauge Dependence of Spontaneous Symmetry Breaking in Gauge
  Theories}.
\newblock {\em Nucl. Phys.}, B101:173--188, 1975.

\bibitem{Gradshteyn1980}
I.~S. Gradshteyn and I.~M. Ryzhik.
\newblock {\em Tables of integrals, series and products (4th Ed)}.
\newblock Academic Press, Inc. New York, 1980.

\bibitem{Arnone:2001iy}
Stefano Arnone, Yuri~A. Kubyshin, Tim~R. Morris, and John~F. Tighe.
\newblock {Gauge invariant regularization via SU($N|N$)}.
\newblock {\em Int. J. Mod. Phys.}, A17:2283--2330, 2002, hep-th/0106258.

\bibitem{Matt1}
Matthew~P Kellett and Tim~R. Morris.
\newblock in preparation.
\newblock 2017.

\bibitem{Hollands:2002yb}
Stefan Hollands and Robert~M. Wald.
\newblock {An Alternative to inflation}.
\newblock {\em Gen. Rel. Grav.}, 34:2043--2055, 2002, gr-qc/0205058.

\bibitem{Kofman:2002cj}
Lev Kofman, Andrei~D. Linde, and Viatcheslav~F. Mukhanov.
\newblock {Inflationary theory and alternative cosmology}.
\newblock {\em JHEP}, 10:057, 2002, hep-th/0206088.

\bibitem{Hollands:2002xi}
Stefan Hollands and Robert~M. Wald.
\newblock {Comment on inflation and alternative cosmology}.
\newblock 2002, hep-th/0210001.

\bibitem{Carroll:2010aj}
Sean~M. Carroll and Heywood Tam.
\newblock {Unitary Evolution and Cosmological Fine-Tuning}.
\newblock 2010, 1007.1417.

\bibitem{Hasenfratz:1989pk}
P.~Hasenfratz and H.~Leutwyler.
\newblock {Goldstone Boson Related Finite Size Effects in Field Theory and
  Critical Phenomena With O($N$) Symmetry}.
\newblock {\em Nucl. Phys.}, B343:241--284, 1990.

\bibitem{Green:1982sw}
Michael~B. Green, John~H. Schwarz, and Lars Brink.
\newblock {N=4 Yang-Mills and N=8 Supergravity as Limits of String Theories}.
\newblock {\em Nucl. Phys.}, B198:474--492, 1982.

\bibitem{Kikkawa:1984cp}
Keiji Kikkawa and Masami Yamasaki.
\newblock {Casimir Effects in Superstring Theories}.
\newblock {\em Phys. Lett.}, B149:357--360, 1984.

\bibitem{Sakai:1985cs}
N.~Sakai and I.~Senda.
\newblock {Vacuum Energies of String Compactified on Torus}.
\newblock {\em Prog. Theor. Phys.}, 75:692, 1986.
\newblock [Erratum: Prog. Theor. Phys.77,773(1987)].

\bibitem{Feldbrugge:2017mbc}
Job Feldbrugge, Jean-Luc Lehners, and Neil Turok.
\newblock {No Rescue for the No Boundary Proposal}.
\newblock 2017, 1708.05104.

\bibitem{Gupta:1954zz}
Suraj~N. Gupta.
\newblock {Gravitation and Electromagnetism}.
\newblock {\em Phys. Rev.}, 96:1683--1685, 1954.

\bibitem{Kraichnan:1955zz}
Robert~H. Kraichnan.
\newblock {Special-Relativistic Derivation of Generally Covariant Gravitation
  Theory}.
\newblock {\em Phys. Rev.}, 98:1118--1122, 1955.

\bibitem{Feynman:1996kb}
R.~P. Feynman.
\newblock {\em {Feynman lectures on gravitation}}.
\newblock 1996.

\bibitem{Weinberg:1965rz}
Steven Weinberg.
\newblock {Photons and gravitons in perturbation theory: Derivation of
  Maxwell's and Einstein's equations}.
\newblock {\em Phys. Rev.}, 138:B988--B1002, 1965.

\bibitem{Ogievetsky:1965}
V.~I. Ogievetsky and I.~V. Polubarinov.
\newblock Interacting field of spin 2 and the einstein equations.
\newblock {\em Annals Phys.}, 35:167, 1965.

\bibitem{Wyss:1965}
W~Wyss.
\newblock Zur unizit\"at der gravitationstheorie.
\newblock {\em Helvetica Physica Acta}, 38:469, 1965.

\bibitem{Deser:1969wk}
Stanley Deser.
\newblock {Selfinteraction and gauge invariance}.
\newblock {\em Gen. Rel. Grav.}, 1:9--18, 1970, gr-qc/0411023.

\bibitem{Boulware:1974sr}
David~G. Boulware and Stanley Deser.
\newblock {Classical General Relativity Derived from Quantum Gravity}.
\newblock {\em Annals Phys.}, 89:193, 1975.

\bibitem{Fang:1978rc}
J.~Fang and C.~Fronsdal.
\newblock {Deformation of Gauge Groups. Gravitation}.
\newblock {\em J. Math. Phys.}, 20:2264--2271, 1979.

\bibitem{Wald:1986bj}
Robert~M. Wald.
\newblock {Spin-2 Fields and General Covariance}.
\newblock {\em Phys. Rev.}, D33:3613, 1986.

\bibitem{Boulanger:2000rq}
Nicolas Boulanger, Thibault Damour, Leonardo Gualtieri, and Marc Henneaux.
\newblock {Inconsistency of interacting, multigraviton theories}.
\newblock {\em Nucl. Phys.}, B597:127--171, 2001, hep-th/0007220.

\bibitem{Mohr:2015ccw}
Peter~J. Mohr, David~B. Newell, and Barry~N. Taylor.
\newblock {CODATA Recommended Values of the Fundamental Physical Constants:
  2014}.
\newblock {\em Rev. Mod. Phys.}, 88(3):035009, 2016, 1507.07956.

\bibitem{Morris:2016nda}
Tim~R. Morris and Anthony W.~H. Preston.
\newblock {Manifestly diffeomorphism invariant classical Exact Renormalization
  Group}.
\newblock {\em JHEP}, 06:012, 2016, 1602.08993.

\bibitem{Stelle:1976gc}
K.~S. Stelle.
\newblock {Renormalization of Higher Derivative Quantum Gravity}.
\newblock {\em Phys. Rev.}, D16:953--969, 1977.

\bibitem{Bergere:1975tr}
M.~C. Bergere and Yuk-Ming~P. Lam.
\newblock {Equivalence Theorem and Faddeev-Popov Ghosts}.
\newblock {\em Phys. Rev.}, D13:3247--3255, 1976.

\bibitem{Itzykson:1980rh}
C.~Itzykson and J.~B. Zuber.
\newblock {\em {Quantum Field Theory}}.
\newblock International Series In Pure and Applied Physics. McGraw-Hill, New
  York, 1980.

\end{thebibliography}

\end{document}